\documentclass[smallextended]{svjour3}       
\usepackage{fullpage, amsmath, amssymb, array, bbm}
\usepackage{natbib,enumerate,gensymb,lineno}

\smartqed  
\usepackage{graphicx}
 \journalname{for arXiv}
\begin{document}

\title{Variable Synaptic Strengths Controls the Firing Rate Distribution in Feedforward Neural Networks}


\titlerunning{Firing Rate Heterogeneity}        

\author{Cheng Ly$^{1,}$*
	\and 
        Gary Marsat$^{2,}$*\thanks{* Equal contribution.  Correspondences about experimental data should be addressed to gary.marsat@wvu.edu; about computational model to: CLy@vcu.edu} 
}


\institute{1. C. Ly \at
	     Department of Statistical Sciences and Operations Research \\
	     Virginia Commonwealth University \\
              Richmond, Virginia 23284-3083 USA \\
              Tel.: +1(804) 828-5842\\
              Fax: +(804) 828-8785\\
              \email{CLy@vcu.edu}           
	\and
		2. G. Marsat \at
	     Department of Biology \\
	     West Virginia University \\
              Morgantown, West Virginia 26506 USA \\
              Tel.: +1(304) 293-2126\\
              \email{gary.marsat@wvu.edu}   
}

\date{Received: date / Accepted: date}

\maketitle

\begin{abstract}

Heterogeneity of firing rate statistics is known to have severe consequences on neural coding.  
Recent experimental recordings in weakly electric fish indicate that the distribution-width of superficial pyramidal cell firing rates (trial- and time-averaged) 
in the electrosensory lateral line lobe (ELL) depends on the stimulus, and 
also that network inputs can mediate changes in the firing rate distribution across the population.  
We previously developed theoretical methods to understand how two attributes (synaptic and intrinsic heterogeneity) interact and alter the firing rate distribution in a population of integrate-and-fire neurons with 
random recurrent coupling.  Inspired by our experimental data, we extend these theoretical results to a delayed feedforward spiking network that qualitatively capture the changes of firing rate heterogeneity 
observed in in-vivo recordings.  
We demonstrate how heterogeneous neural attributes alter firing rate heterogeneity, accounting for the effect with various sensory stimuli.  
The model predicts how the strength of the effective network connectivity is related to intrinsic heterogeneity in such delayed feedforward networks: the strength of the feedforward input 
is positively correlated with 
excitability (threshold value for spiking) with low firing rate heterogeneity and is negatively correlated with excitability with high firing rate heterogeneity.  We also show how our theory 
can be use to predict effective neural architecture.  We demonstrate that neural attributes do not interact in a simple manner but rather in a complex stimulus-dependent 
fashion to control neural heterogeneity and discuss how it can ultimately shape population codes.  

\end{abstract}

\section{Introduction}
\label{intro}

The mechanisms and features of neural networks that enable efficient sensory coding is an important topic with many advances stemming from a combination of experiments and computational modeling.  
In a population of neurons, understanding how sensory signals are encoded and transmitted to higher areas of the brain is especially challenging given that neurons, even in early stages of processing, are known to be 
stochastic and have heterogeneous attributes.  The structure and distribution of this heterogeneity will thus determine how the population of neurons jointly encode a stimulus.  
Firing rate heterogeneity has been shown to have consequences on 
neural coding in the olfactory bulb \citep{padmanabhan10,tripathy13}, and in models of the visual system with diverse orientation tuning curves \citep{shamir06,chelaru08}, and a variety of other systems 
\citep{georgopoulos86,marsat10,ahn2014}.  
In general, the firing rate heterogeneity is known to affect important information-theoretic measures of coding such as the Fisher information and mutual information \citep{kaybook}.  This quantity is a significant 
measure for systems that code signals based on rate, or the total number of spikes.  
Although we do not focus on the potential impact of higher order spiking statistics, firing rate heterogeneity has a direct impact on the way the population encodes sensory signals.  
The primary focus of this paper is on the firing rate heterogeneity (distribution) of pyramidal cells in a delayed feedforward network 
that is motivated by our data in the Electrosensory Lateral Line lobe (ELL) of apteronotids weakly electric fish.

The weakly electric fish is a well established model of sensory processing that continues to provide powerful insight into the neural dynamic of sensory coding. The ELL is particularly well understood \citep{maler09}, 
it is the sole gateway from the peripheral electrosensory receptors to higher sensory areas. The principle cells of this network -- pyramidal cells -- receive direct input from receptors and project to the next level of 
sensory processing. A subset of pyramidal cells (the so-called superficial and intermediate ON-cells on the lateral segment) is the focus of the data we present 
because they receive both the direct feedforward inputs and a large set of inputs from parallel fibers projections. 
Parallel fibers originate for the caudal lobe of the cerebellum which is driven by input from another subset of ELL pyramidal cells, deep pyramidal cells. Therefore, although this parallel fiber input to superficial pyramidal cells is 
traditionally described as feedback, it is an open loop configuration and thus can be regarded as an indirect delayed feedforward input. Pyramidal cell response heterogeneity (even within the superficial subset) has been shown to 
be important for coding of different types of natural signals \citep{marsat10,marsat12}. 
Network dynamic, in particular parallel fiber input, can influence heterogeneity and correlation among responses \citep{litwin12,simmonds15}. 
We present {\it in vivo} data from our lab (see Fig. 1{\bf B}--{\bf C}) that indicates that population firing rate heterogeneity can be modulated in a stimulus-dependent manner so as to shape the population code, 
an observation consistent with our prior results \citep{marsat14}.  Specifically, low frequency  
stimuli typical of male-male aggressive interaction elicit population responses with low firing rate heterogeneity whereas high frequency stimuli typical of male-female interactions and 
courtship lead to higher heterogeneity. Low or 
high frequency sinusoidal amplitude modulations of the fish's electric field are present during any interaction with conspecific. These sine waves are thus the natural background signal that set the neural 
dynamic in a specific 
mode, thus influencing the processing of transient communication signals. This observed change in heterogeneity of the population raises a question: how can a single population of cells change its response heterogeneity instead 
of the heterogeneity being a fixed attribute of the population? A simple answer would be that the inputs to the cells change as a function of the stimulus; another answer is that input heterogeneity could interact with intrinsic 
properties to produce complex changes in response heterogeneity.  
Our goal in this paper is to investigate theoretically what mechanism of a plausible feedforward network 
of the ELL system can lead to such phenomenon.


Here we focus on the firing rate averaged in time and over trials (i.e., depicted by the arrows in Fig. \ref{fig:setup}{\bf B}\&{\bf C}), with {\it heterogeneity} referring to the different firing rate distributions across the 
population, measured by the standard deviation across different cell rates.  We consider two sources of firing rate heterogeneity: intrinsic and network. 
Many intrinsic factors influence the firing rate of a cell such as ion channel composition or cell morphology. Arguably, the most central parameter that dictates 
firing rate is the threshold of the cells since low threshold will directly cause high firing rates and vice versa. Threshold heterogeneity has been shown experimentally in cortical cells \citep{azouz00} and has 
crucial effects in the electrosensory system \citep{middleton09} and others \citep{priebe08}.  
We therefore use threshold as the source of intrinsic heterogeneity across our population of cells. 
Note that a cell's threshold is itself dictated by a variety of factors but it is not our goal to detail the underlying molecular dynamic at play. Network heterogeneity refers to any aspect of the network inputs that can influence the 
cell's firing. Here again we focus on the simplest parameter of network input affecting a cell's firing: input strength \citep{marder06,chelaru08}. 
Input strength is determined by many physiological parameters: presynaptic firing rate, PSP size \citep{bremaud_07} for each 
presynaptic spike or number of inputs \citep{parker2003variable,oswald09} to name only a few. We do not distinguish here between these different factors. We seek to determine how these two sources of firing rate heterogeneity interact.  
Thus, we adapt and apply a previously developed theoretical framework \citep{Ly_15} to a delayed feedforward spiking network model of the ELL electrosensory pathway.  The model can qualitatively capture the 
different firing rate heterogeneity (measured by standard deviation) 
depending on different stimuli, and enables an experimental prediction about how the effective network connectivity is related to intrinsic heterogeneity.  Specifically, the fitted model along with our theory 
predicts that, when electrosensory stimuli is low frequency (a signature of same-sex interactions), 
target pyramidal cells that are less excitable (higher spike thresholds) have relatively {\it stronger} excitatory and inhibitory presynaptic input and cells that are more excitable (lower spike thresholds) have 
{\it weaker} excitatory and inhibitory presynaptic input.  When the stimulus is high frequency (a signature of opposite-sex courtship), the opposite happens: 
target pyramidal cells that are less excitable (higher spike thresholds) have relatively {\it weaker} excitatory and inhibitory presynaptic input and cells that are more excitable (lower spike thresholds) have 
{\it stronger} excitatory and inhibitory presynaptic input.  

We further demonstrate the value of the theory by showing how the firing rate standard deviations from our data 
can be captured with a delayed feedforward network model with fixed synaptic input strengths and different network architecture, as opposed to changing the strengths in the prescribed way.  
Our work demonstrates how theoretical 
analysis can be used to elucidate the interactions of neural attributes with various stimuli, and to investigate 
how presynaptic inputs can shape the network firing statistics.  
Given the widespread nature of feedforward pathways in the nervous system \citep{berman99,ferster00,bruno02,pouille01,bastian04} and the generic structure and parameters in our model, 
our results might characterize a general mechanism applicable to a variety of systems.

\section{Methods}
\label{sec:1}

\subsection{Delayed feedforward network model}

Our {\it in vivo} experimental data from pyramidal cells of weakly electric fish's hindbrain, presented in section \ref{sec:electr}, motivates our theoretical application.  
The feedforward spiking model we present mimics relevant features of the weakly electric fish system, but is also a rather generic feedforward model so that our theoretical 
results might be operative in other systems.

The population of interest consists of hindbrain pyramidal cells (only superficial and intermediate pyramidal ON-cells of the lateral segment of the ELL are recorded) 
receiving afferent sinusoidal input (\ref{fig:setup}{\bf A}) and network input via the parallel fibers from granule cells (often termed `feedback' in the ELL of electric fish 
hindbrain even though it an open-loop and thus can be considered as delayed feedforward input).  
We model the parallel fiber input with an equal number of excitatory ({\bf E}) and inhibitory ({\bf I}) inputs; the E cells are driven by afferent sinusoidal input while the I cells receive input from these E cells.  
The granule cell input (aforementioned E cells) and local interneuron input (aforementioned I cells) to the superficial pyramidal cells are delayed by $\mathcal{O}(10)\,$ms to mimic the ELL pathway. 
This configuration captures the essence of parallel fiber input in steady state \citep{bol11,maler07}.  All cells are 
modeled as leaky integrate-and-fire (LIF) point neurons.  Note that we have chosen to exclude the specific attributes known in the ELL (e.g., bursting mechanisms, synaptic plasticity, etc.) to have a rather general 
feedforward network and mechanisms that do not rely on the particularities of this specific system.  The general structures of such networks are common to many pathways and areas of the nervous system 
(visual system \citep{ferster00}, somatosensory system \citep{bruno02}, hippocampus \citep{pouille01}, electrosensory system \citep{berman99,bastian04}, etc.).  
The intrinsic heterogeneity and synaptic variability are modeled simply by two parameters that are allowed to vary among the pyramidal neurons. 

The equations for the (target) pyramidal neurons indexed by $j\in\{1,2,\dots,N\}$ are:
\begin{eqnarray}\label{e_lif}
	\tau_m \frac{d v_j}{dt} & = & I_{aff}(t)-v_j-q_j g_{i,j}(t-\tau_{del})(v_j-\mathcal{E}_I)-q_j g_{e,j}(t-\tau_{del})(v_j-\mathcal{E}_E)+\sigma_P\eta_j(t) \nonumber \\
	v_j(t^*) & \geq & \theta_j \hbox{(refractory period)} \Rightarrow v_j(t^*+\tau_{ref})=0 \nonumber \\
	\tau_n \frac{d \eta_j}{dt} & = & -\eta_j+\sqrt{\tau_n}\xi_j(t)  \nonumber \\
	I_{aff}(t) &:=& \lceil I_0+\mathcal{A}\sin(2\pi\phi t) \rceil^+
\end{eqnarray}
where the leak, inhibitory and excitatory reversal potentials are 0, $\mathcal{E}_I$, and $\mathcal{E}_E$, respectively with $\mathcal{E}_I<0<\mathcal{E}_E$, 
(the voltage is scaled to be dimensionless so that a leak/resting value of -65\,mV maps to 0 and a threshold voltage of -55\,mV maps to 1 (the average threshold), see Table \ref{tab:1} for other parameters), 
and $\xi_j(t)$ are uncorrelated (in time and across neurons) white noise processes with 0 mean and unit variance.  
Sinusoidal stimuli are a realistic model of the signal the electric fish is exposed to during interactions with conspecific (i.e., the so-called beat amplitude modulation \citep{maler07}).  
We model the afferent input $I_{aff}(t)$ as a positively rectified sinusoid that includes nonlinearities 
(e.g. saturation or rectification) -- here, $\lceil x \rceil^+=0$ if $x\leq 0$; $\lceil x \rceil^+=x$ if $x>0$.  The stimuli are predominately linear \citep{gussin07}.  
The second line in the equations describes the refractory period at spike time $t^*$: when the neuron's voltage crosses 
threshold $\theta_j$ ({\bf intrinsic heterogeneity}), 
the neuron goes into a refractory period for $\tau_{ref}$ where the voltage is undefined\footnote{In refractory, the other variables are governed by their ODEs}, 
after which we set the neuron's voltage to 0. 

The conductance variables ($g_{i,j}(t-\tau_{del})$ and $g_{e,j}(t-\tau_{del})$) are determined by the delayed feedforward input (equations to follow) and are \underline{both} scaled by a factor $q_j$ that is specific to the $j^{th}$ neuron 
because the inputs are disynaptic \citep{maler07}.  
The parameter 
$q_j$ introduces {\bf synaptic variability} that is loosely motivated by recent results by \citet{xue14}, who found that pyramidal neurons receive relatively similar proportions of excitation and inhibition in 
layer 2/3 of mammalian visual cortex (i.e., some cells receive more E/I while some cells receive less E/I).  This type of synaptic variability has been used in 
other models to study heterogeneity \citep{chelaru08}.  
Note that synaptic variability can be distinct for variability in the structure of the network (i.e. number of connections) but in our case, it does not need to be so. 
Here, the synaptic variability $q_j$ represent any aspects of the network input that lead to the variability in its strength from cell to cell.

The equations for the cells modeling the delayed feedforward network inputs are similar in form but have different parameter values, and their activity determines the synaptic conductance values in 
the aforementioned population.  We use a simple model of the effective feedforward inputs to the superficial pyramidal neurons in the ELL system \citep{maler09,chacron05}, see Figure \ref{fig:setup}{\bf A}.  
There are $2N_f$ cells in the delayed feedforward population, with equal numbers of excitatory (i.e., granule cells) and inhibitory (i.e., local interneurons) cells: $N_f$.  First, the equations for the granule cells 
that only provide excitatory input are (for $l\in\{1,2,\dots,N_f\}$):
\begin{eqnarray}\label{f_lif}
	\tau_m \frac{d v_l}{dt} & = & I_{aff}(t)-v_l+\sigma_F\eta_l(t) \nonumber \\
	v_l(t^*) & \geq & 1 \hbox{(refractory period)} \Rightarrow v_l(t^*+\tau_{fref})=0 \nonumber \\
	\tau_n \frac{d \eta_l}{dt} & = & -\eta_l+\sqrt{\tau_n}\xi_l(t)  \nonumber \\
	\tau_d\frac{d G_l}{dt} &=& -G_l + A_l  \nonumber \\
	\tau_r\frac{d A_l}{dt} &=& -A_l + \tau_r \alpha \sum_{k} \delta(t-t_k) \nonumber \\
	g_{e,j}(t) &=& s_e \sum_{l'\in\{\hbox{granule cells}\}} W_{j,l'} G_{l'}(t) 
\end{eqnarray}
Here $t_k$ are the spike times of the particular granule cell, and $W_{j,l'}\in\{0,1\}$ is an $N\times N_f$ matrix that specifies the network connectivity.

The equations for the local interneurons are similar in form to the granule cells (Eq. \eqref{f_lif}) with many of the same parameters, except that these cells do not 
receive direct sinusoidal input but rather receive excitatory inputs from the granule cells.  For $n\in\{1,2,\dots,N_f\}$, we have:
\begin{eqnarray}\label{lif_inter}
	\tau_m \frac{d v_n}{dt} & = & -v_n- g_{EE}(t)(v_n-\mathcal{E}_E)+\sigma_F\eta_n(t) \nonumber \\
	v_n(t^*) & \geq & 1 \hbox{ (refractory period)} \Rightarrow v_n(t^*+\tau_{ref})=0 \nonumber \\
	\tau_n \frac{d \eta_n}{dt} & = & -\eta_n+\sqrt{\tau_n}\xi_n(t)   \nonumber \\
	g_{EE}(t) & =& s_{EE} \frac{1}{N_f}  \sum_{l'=1}^{N_f} G_{l'}(t) \nonumber \\
	\tau_d\frac{d G_n}{dt} &=& -G_n + A_l  \nonumber \\
	\tau_r\frac{d A_n}{dt} &=& -A_n + \tau_r \alpha \sum_{k} \delta(t-t_k) \nonumber \\
	g_{i,j}(t) &=&  s_i \sum_{n'\in\{\hbox{I cells}\}} W_{j,n'} G_{n'}(t) 
\end{eqnarray}
Here $t_k$ are the spike times of the particular local interneuron.  
The response of these cells provide the feedforward inhibitory input in equation \eqref{e_lif}.  The parameter values are in Table \ref{tab:1}, or will vary and be specified later.

There are various levels of modeling of the ELL circuit (multicompartment \citep{doiron01} to single compartment \citep{litwin12,mejias13} models).  
The delayed feedforward input in total consists of both direct excitatory inputs and inhibition via local interneurons.  
We explicitly add an extra synapse to more faithfully model the anatomy of the ELL system: here, the 
local interneurons (Eq. \eqref{lif_inter}) provide the synaptic inhibitory input while the granule cells (Eq. \eqref{f_lif}) only provide excitatory input to both the superficial/intermediate pyramidal cells and the local interneurons (see Fig. \ref{fig:cycl}{\bf A}).  
However, models of this system by others \citep{doiron03,chacron05,litwin12,mejias13,simmonds15} 
do not model the delayed feedforward inhibition with a differential equation model of the intermediate local interneurons.

The mean firing rate $\nu_j$ is defined by:
\begin{equation}\label{frt_defn}
	\nu_j := \frac{\hbox{Number of spikes}}{\hbox{Total time}}
\end{equation}
The mean firing rate is a common statistical quantity of interest and is, among other spike metrics, known to have important implications for encoding signals \citep{kaybook}.

Initially, the delayed feedforward network is randomly connected to the superficial pyramidal neurons with a 20\% connection probability (Erd\H{o}s-R\'enyi graph).  
We also set $s_e=2.3/(0.2*N_f)=0.0575$ and $s_i=1/(0.2*N_f)=0.025$ in all of the figures, including when the connectivity is no longer Erd\H{o}s-R\'enyi.

\begin{table}
\caption{Parameter values for delayed feedforward network}
\label{tab:1}       
\begin{tabular}{ll}
\hline\noalign{\smallskip}
{\bf Parameter} & Value \\ \hline 
$N$ & 1,000 \\
$\tau_m$ & 10\,ms \\
$\tau_{ref}$ &  1\,ms \\
$\mathcal{E}_I$  &  -0.5 \\
$\mathcal{E}_E$ &  6.5\\
$\sigma_P$ & 	0.75 \\
$\tau_{del}$ &   20\,ms	\\
$\tau_n$ &    	5\,ms \\
$I_0$ & 	0.35 ($\phi=5$\,Hz), 0.4 ($\phi=120$\,Hz) \\
$\mathcal{A}$ & 	0.35 ($\phi=5$\,Hz), 0.55 ($\phi=120$\,Hz) \\	 \hline
$N_f$ & 	100 \\
$s_{EE}$ & 2.7 \\
$\tau_{fref}$ & 0.5\,ms		\\
$\sigma_F$ &	1\\
$\tau_d$ & 10\,ms 		\\
$\tau_r$ & 2\,ms  \\
$\alpha$ &  2 
\end{tabular}

\end{table}

Response for the intact network 
that contains delayed feedforward network inputs from the granule cells is shown in black (Fig. \ref{fig:setup}{\bf B}).  This network input can be blocked pharmacologically so that the pyramidal cells only receive direct afferent 
input (see section \ref{sec:electr}). 
A plausible model should take the effect of granule cell network input into account and replicate the effect of blocking this delayed feedforward input.  
This model and our previously developed theory \citep{Ly_15} enables the structure of the synaptic variability to dramatically affect the firing 
rate distribution.  


\subsubsection{Distributions of the intrinsic heterogeneity and synaptic variability}

The two parameters $(q_j,\theta_j)$ are varied to evaluate their effect on firing rate distributions.  The means of both $\vec{q}$ and $\vec{\theta}$ are set to 1, and the 
parameters $\sigma_q$ and $\sigma_\theta$ quantify the level of the synaptic variability and intrinsic heterogeneity, in the following way:
\begin{eqnarray}\label{het_sig1}
	\vec{q} &\sim& 1+\sigma_q*(\mathcal{U}-0.5) \\
	\vec{\theta} &\sim& e^{\mathcal{N}} \label{het_sig2}
\end{eqnarray}
where $\mathcal{U}$ is the uniform distribution on $[0,1]$, and $\mathcal{N}$ is normal distribution with mean $-\sigma^2_\theta/2$ and standard deviation $\sigma_\theta$, so that $\vec{\theta}$ has a 
log-normal distribution with mean 1 and variance: $e^{\sigma_\theta^2}-1$.  Throughout this 
paper, we set $\sigma_\theta=0.1$ and $\sigma_q=1$, which can give a wide distribution of firing rates.

\subsubsection{Changing the correlation between intrinsic heterogeneity and synaptic variability}\label{sec:corr_sig}

A key way to change the firing rate heterogeneity, where the overall level of heterogeneity of $\vec{q}$ and $\vec{\theta}$ are approximately the same, is by setting the correlation between $\vec{q}$ and $\vec{\theta}$ to 
a particular value.  Given the vectors $\vec{q}$ and $\vec{\theta}$, we fix $\vec{q}$ to the same values but transform $\vec{\theta}$ so that the Pearson's correlation coefficient is 
$\varrho\in(-1,1)$ in such a way that the transformed vector has the same mean and variance as original $\vec{\theta}$.  The methods used to accomplish this were previously described in the Appendix of \citet{Ly_15}.  
In the following figures, the mean firing rates do not vary much as the correlation of these two parameters vary, enabling our theoretical results to focus on firing rate heterogeneity.


\subsection{Electrophysiology}\label{sec:electr}

Surgical and experimental procedure follow established methods \citep{bol11,bastian86,chacron05}.  
{\it A. leptorhynchus} adult males or females were anesthetized with tricaine methanesulfonate and respirated during surgery.  
The skull above the ELL was removed after local anesthetic was applied to the wound.  The skull was glued to a post for stability.  General anesthesia was stopped after 
the fish was immobilized with an injection of turbocurare.  Experiments were conducted in a large tank (35 cm $\times$ 35 cm $\times$ 20 cm) 
with home-tank conditioned water (26$\degree$C and a conductivity of 250 $\mu$S). In vivo, single-unit recordings from superficial and intermediate (spontaneous firing rate $<20$\,Hz) 
ON-cells of the lateral segment were performed using metal-filled extracellular electrodes \citep{frankBecker64}.  
Pyramidal cells can easily be located by the anatomy of the ELL and overlying cerebellum as well as by their response properties \citep{maler91,bastian84}.  
All experiments and protocols were approved by the Institutional Animal Care and use Committee.  

Superficial and intermediate pyramidal cells have large apical dendrites where feedback inputs synapse. Deep pyramidal cells do not receive feedback. Therefore, we could ascertain that the 
cells we recorded from are not deep cells by verifying that the feedback block had an influence on responses (see below). 
Furthermore, superficial and intermediate have low spontaneous firing rate ($<25$\,Hz; \citep{bastian01}) thereby allowing a rapid identification of cells type.  
Our data set includes 15 superficial/intermediate pyramidal cells stimulated with continuous AM (pure) sinewaves at two frequencies: 5\,Hz (low) and 120\,Hz (high).  
Stimuli are applied in a spatially global configuration with a pair of electrodes positioned on opposite sides of the tank. 
This spatial configuration drives both the afferent inputs to the pyramidal cells but also the delayed feedforward network. 
The delayed feedforward input can be blocked by injecting lidocaine in the axon bundle that innervates the caudal cerebellum. 
The injected lidocaine does not directly affect pyramidal cells.  This was verified by comparing responses to step stimuli presented in between each sinusoidal stimuli and delivered 
through a small local dipole. The small local dipole drive the direct feedforward inputs but not the delayed feedfoward inputs via parallel fiber, we used recording only if the lidocaine injection 
did not affect the response to our local step stimuli.  
We also verified that our injection effectively blocked the feedback pathway by comparing responses to 5\,Hz sine waves presented via the regular global electrodes while the feedback is blocked or not. 
We verified that the firing 
rate was significantly affected either at the peak or at the trough. Based on the mean firing rate over a quarter of a cycle (centered around the phase most affected by feedback) each cell presented a significant difference either 
at the trough (feedback block decreased firing rate) or at the peak (feedback block increased firing rate).  The significance of this changed was tested with a paired t-test (significance level p$<0.05$).  
The details of the stimulation, recording and lidocaine block are the same as those described in \citet{marsat2012prep} and also used in other studies \citep{bastian86,chacron05}.

The delayed feedforward input is usually described as a feedback input.  
In our experiment, an immobilized fish, this feedback pathway would not receive significant proprioceptive inputs and, in this species, there is no efferent copy input to the feedback pathway.  
It is therefore an appropriate simplification to describe this feedback input as delayed feedforward since it is essentially an open-loop pathway that does not receive a strong drive from the neuron it targets.

\section{Results}

\begin{figure}
\centering
	\includegraphics[width=\columnwidth]{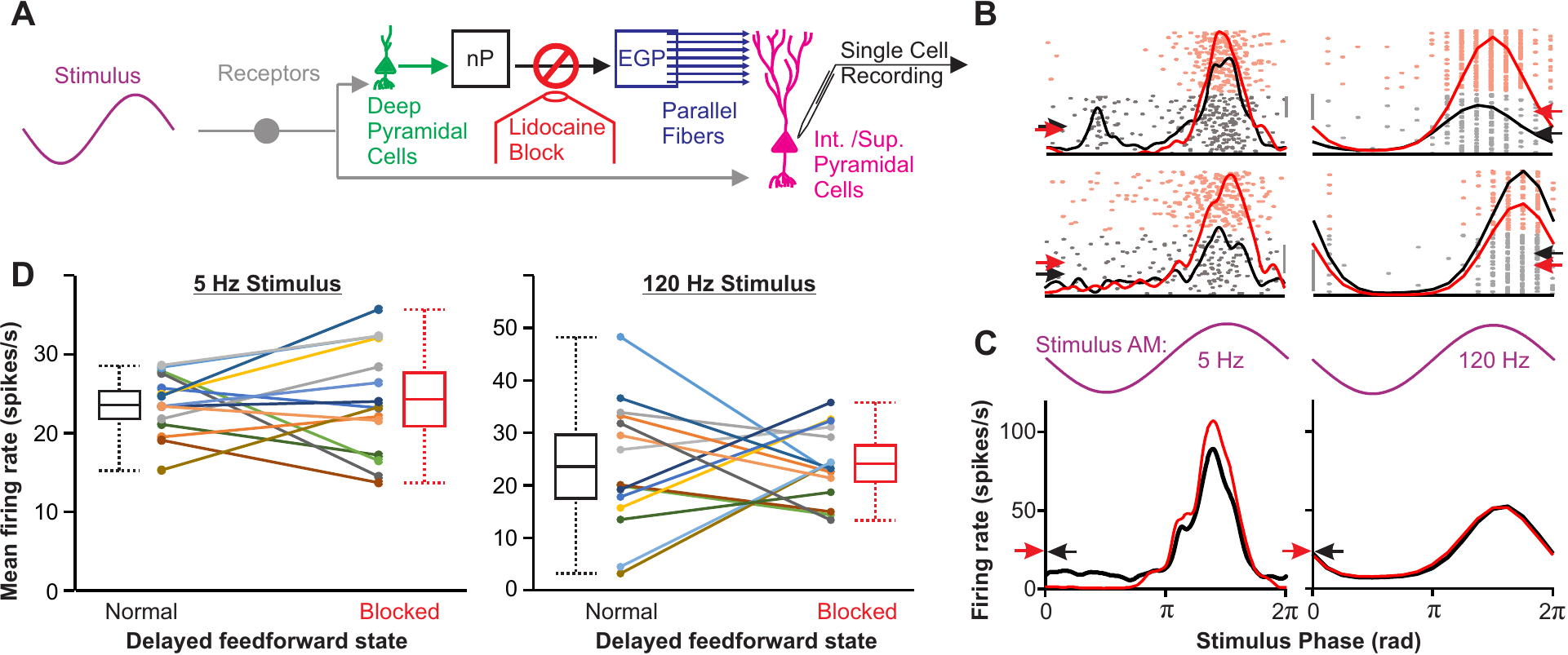}
\caption{ In the ELL of weakly electric fish, firing rate heterogeneity can be increased or decreased by network inputs. 
{\bf A}) Schematic of the relevant circuitry of the experimental system. The sinusoidal stimulus' amplitude is encoded by p-unit electroreceptor which project to pyramidal cells of the ELL.  
Deep pyramidal cells are the first step of the delayed feedforward pathway which goes through the nucleus praeeminentialis dorsalis (nP), the eminential granularis posterior (EGP) and onto the 
apical dendrites of superficial and intermediate pyramidal cells of the ELL via granular cell's parallel fibers. This delayed feedforward pathway can be blocked by lidocaine injection in the fiber bundle that 
provide inputs to the EGP. {\bf B}) Example responses of 2 individual pyramidal cells for each stimulus frequency. The average instantaneous firing rate was calculated by 
Gaussian-convolution (3\,ms width) of the binarized spike train and then averaging across cycles of the 15+ sec stimuli.  A raster plot of these responses is displayed in the background. 
Red is used for the responses while the delayed feedforward is blocked; black for responses while it is intact.  Arrows show the mean firing rates for the neurons for this stimulus and the gray scale bars 
near the central edge of the plots represent firing rates changes of 20\,Hz. These examples show that the mean firing rates of the neurons can be affected one way or another.  
Note that, raster plots have a sampling rate of 2\,kHz and thus, for 120\,Hz sine waves, the raster plot display spike in bins of 1/17th of a cycle.  
{\bf C}) Mean instantaneous firing rate for a population of pyramidal cells responding to the 2 sinusoidal stimuli. The instantaneous firing rates of the cycle-histogram shows that the 
delayed feedforward affects the depth of the modulation of the response to low frequency but not to high frequency stimuli (compare the red and black curves). 
In both cases the mean firing rate (averaged across time and across neurons) is not dependent on the stimulus frequency or the delayed feedforward block. 
{\bf D}) Changes in mean firing rate distribution across the population of cells for low and high frequency responses when the delayed feedforward is blocked. 
We display the mean firing rate when the delayed feedforward is blocked (right side of the plots) or not (left side of the plots) with two dots of a different color for each neurons linked together by a line. 
The slope of the line clearly shows that the block cause an increase in firing rate for some neurons and a decrease in others. 
The box plots show the statistics for this population: mean (central line) standard deviation (box) and range (dotted lines). Clearly, the heterogeneity as measured by the variance is 
increased by the block for 5\,Hz stimuli but decreased for 120\,Hz stimuli.  Statistics: the 4 populations of firing are normally distributed (Kolmogorov-Smirnov tests: 0.89$>$p$>$0.86).  
Variance for low frequency varied significantly between blocked and normal condition (F-test p=0.033) and also for high frequency stimuli (p=0.042). 
Mean firing rate of the whole population (across time and across neurons) did not differ between the blocked vs. normal condition 
(paired T-test assuming unequal variance; low frequency, p=0.78; high frequency, p=0.91).  
Our data show significant differences in variance in normal vs. blocked despite our sample size leading to probabilities of finding significant differences of 
0.62 and 0.63 as tested with a post-hoc Power Analysis \citep{faul07g,faul09} (see \citet{thomas97} and \citet{hoenig01}).  
}
\label{fig:setup}       
\end{figure}

The influence of the parallel fiber input onto superficial pyramidal cells (and intermediates) of the ELL has been intensively studied and showed striking effect on how the depth of the sinusoidally modulated 
neural responses is decreased by the input from parallel fibers \citep{bastian86} for low frequency stimuli. This so-called cancellation of low-frequency response can be observed in our data when comparing the
 responses to the low frequency stimulus (Fig. \ref{fig:setup}{\bf B}\&{\bf D}, left columns): firing rate during the trough is increased by the parallel fiber input and the response during the peak is decreased. 
 In average, the parallel fiber input was not found to produce any effect on mean firing rate \citep{bastian86}. Our data confirm this for the population as a whole since the mean firing rate did not change 
 (see box plots in Fig. \ref{fig:setup}{\bf D} and arrows in Fig. \ref{fig:setup}{\bf C}).  However, it was unknown whether the feedback could influence the mean firing rate of individual cells and 
 whether this influence was frequency dependent. Note that although the most striking effect of the parallel fibers on pyramidal cells is its effect on low-frequency responses, the bulk of inputs to 
 granule cells/parallel fiber respond very well to high frequencies. Thus, nothing prevents this pathway from having an influence for a wide variety of stimuli. Here we observed that for both low and high 
 frequency stimuli, the mean firing rate of individual neurons changed when the delayed feedforward pathway was blocked. For a particular cell, the change was sometimes an increase in the mean firing rate, sometimes a decrease. 
 As a result, the mean firing of the population did not change significantly but the standard deviation did (see Fig. \ref{fig:setup} caption for statistics). Most interestingly, 
 the change in heterogeneity (standard deviation) of mean firing rates across the population was different for low frequency stimuli vs high frequency stimuli. For low frequency stimuli, the delayed feedforward 
 input decreased the heterogeneity (i.e., blocking the feedback increased the standard deviation) whereas it increased it for high frequency stimuli. This interesting effect on firing rate heterogeneity cannot be 
 explained by the intrinsic properties of the cells or the direct feedforward input. We determined, with our model, the condition required to replicate the effect, while also taking in account the known fact that the delayed feedforward 
 pathway provides frequency-specific input \citep{bol11}.

In presenting our model that replicate the experimental effect described above, we 
first demonstrate that a delayed feedforward spiking neural network model can capture the population firing rate features exhibited in the experimental 
data, with only 1 major parameter change (sinusoidal frequency) to capture two distinct types of realistic sensory stimuli.  For better quantitative matching, we slightly alter the mean and amplitude of the 
sinusoidal input in the model depending on frequency (see Table \ref{tab:1}), which replicates the frequency response profile of these pyramidal cells. Indeed our data shows that the peak of the instantaneous 
firing is different for low and high frequencies (Fig. \ref{fig:setup}{\bf D}).  
We apply our theory for mean firing rate (trial- and time-averaged) heterogeneity across the population, to the fitted model to 
determine the relationship of the heterogeneous parameters that captures the firing rate heterogeneity in the data.  Finally, we demonstrate the utility of the theory for {\it effective network connectivity} 
with example networks where the probability of connection is structured in such a way to obtain the prescribed correlation between the (heterogeneous) neural attributes derived from the analysis.

\begin{figure}
	\centering 
	\includegraphics[width=.5\columnwidth]{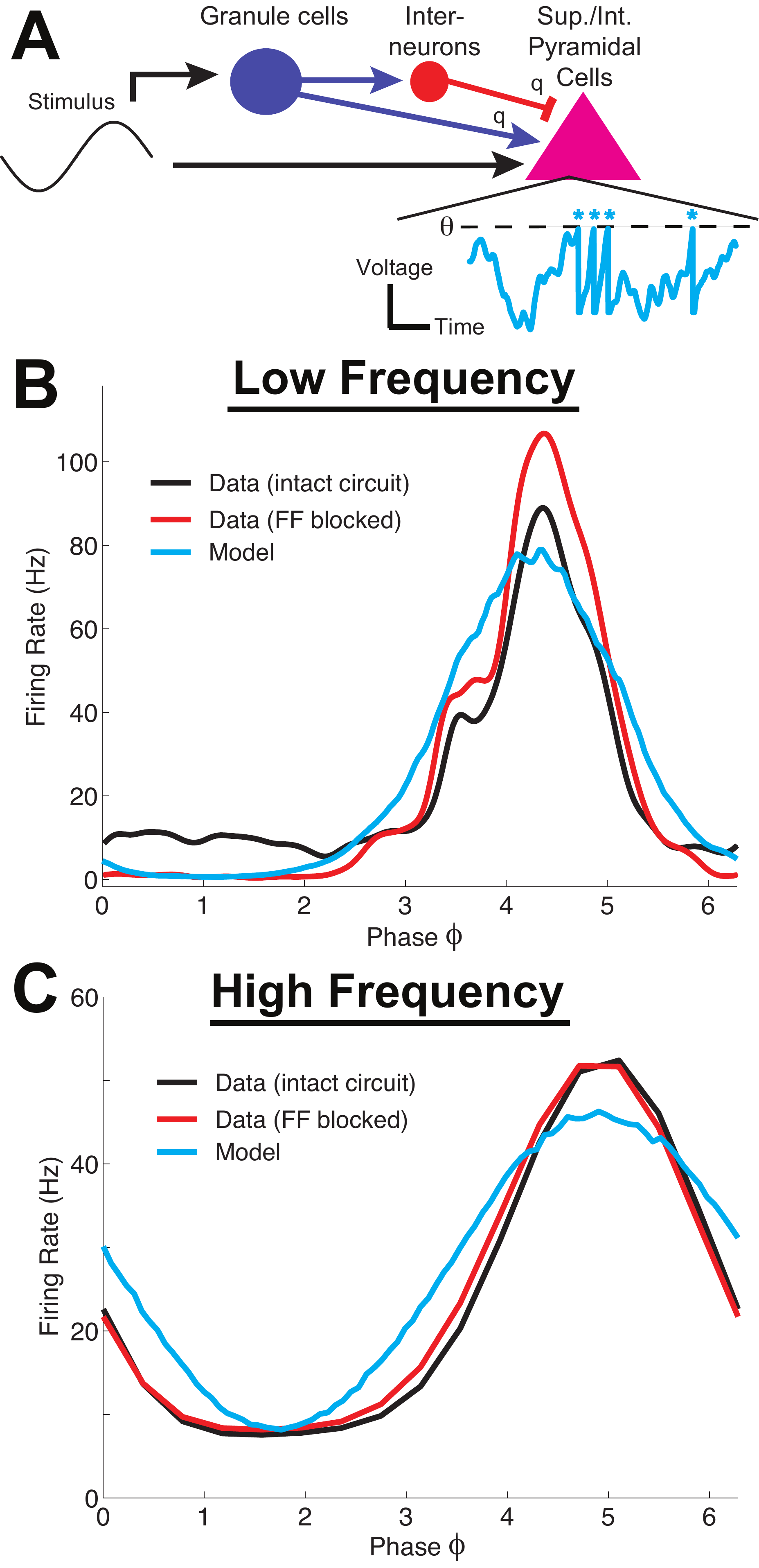}
\caption{Population firing rates of the data and the delayed feedforward model match.  {\bf A}) Model schematic of the delayed feedforward network.  The delayed feedforward model (cyan) not only 
captures the mean population firing rate of the intact circuit (black) well, but also captures the cycle histogram firing rates for both low ({\bf B}) and high ({\bf C}) frequencies.  Also shown is the 
population average cycle histogram from the data when the delayed feedforward input is blocked as in Figure \ref{fig:setup}{\bf B}\&{\bf C}.  
The two frequencies of the afferent stimuli (not shown) in the experiments are 5\,Hz and 120\,Hz, 
which are the same two values used in the model (see $\phi$ in $I_{aff}(t)$ in equations \eqref{e_lif}--\eqref{f_lif}) and is the main difference (see Methods and Table \ref{tab:1}).  
The trial- and time-averaged population firing rates for the low frequency are: 25.4\,Hz (model) and 23.6\,Hz (data); for the high frequency: 26.8\,Hz (model) and 23.6\,Hz (data).  
Here we set $\varrho(\theta,q)=0$ (see {\bf Methods}) in the model.  
}
\label{fig:cycl}       
\end{figure}

\subsection{Adapting the theory for the delayed feedforward network model}

Previously, our analysis of heterogeneous recurrent LIF networks provided insights to how the firing rate distribution changed as the relationship between threshold heterogeneity and 
synaptic variability changed \citep{Ly_15}.  In the weakly electricfish electrosensory system, (sup./int.) pyramidal neurons receive direct feedforward inputs and input that can be qualified as a delayed feedforward input.  
Fortunately, despite the different types of network, the previously developed theory for recurrent networks can be adapted to feedforward networks (see Appendix A).  
In this delayed feedforward network, the resulting simplified PDF equations have less dimensions than a recurrent network.  
The goal of the analysis here is not to accurately capture the time-varying instantaneous firing rates but rather 
demonstrate the principles for how the relationship between heterogeneous attributes changes the variability in {\it mean} firing rates.  

In this system, there is a long history of using LIF neural networks to capture salient experimental results and in using computation/analysis to uncover details of electrosensory processing 
\citep{doiron03,noonan03,bol11,litwin12,mejias13}.  
Given these previous results, it is not surprising that our network model was able to capture the population firing rate dynamics with with two distinct stimuli (see Fig. \ref{fig:cycl}).  
We sought to capture the population firing rate with both low and high frequency inputs by primarily changing one parameter, 
the frequency of the afferent sinusoidal input: $\phi$ (see equations \eqref{e_lif}--\eqref{f_lif}).  To insure quantitative accuracy, we also slightly varied the amplitude $\mathcal{A}$ and mean $I_0$ of the sinusoidal input 
(see Table \ref{tab:1}).  
The two frequencies of the sinusoidal input were obtained directly from the experiments where the afferents were driven at 5\,Hz and 120\,Hz, respectively.  

The average population firing rate as a function of the phase of the input frequency is shown for both low frequency (Fig. \ref{fig:cycl}{\bf B}) and high frequency (Fig. \ref{fig:cycl}{\bf C}).  
The model, which 
has noise (trial-to-trial variability) and the two forms of quenched variability (threshold and synaptic mediated by feedforward inputs), is able to capture the population average response with 
2 distinct afferent stimuli (compare the black and cyan curves).  
Here, the quenched distributions for the threshold heterogeneity $\vec{\theta}$ and synaptic variability $\vec{q}$ were chosen {\it independently} with 
$\varrho=0$ (see equations \eqref{het_sig1}--\eqref{het_sig2} and Methods for parameter values).  
We did not use an optimization routine based on particular algorithms and did not make specific choices about parameters ranges and other 
parameter attributes; rather, we manually varied the noise levels $\sigma_{P/F}$ and other parameters to match the data.  Since the 
purpose of our computational modeling is to illustrate a principle of network dynamics (specifically effective network connectivity and its relationship with intrinsic excitability/threshold), our results 
hold equally well with other sets of parameters we used in these same class of models.  

The two population of presynaptic (feedforward) cells are relatively simple homogeneous LIF models, with firing rates that depend on the details of the afferent input $I_{aff}(t)$.  When $\phi=5\,$Hz, the time-averaged 
firing rates are $\langle r^E(t) \rangle_t=8.5$\,Hz and $\langle r^I(t) \rangle_t=19.4$\,Hz; when $\phi=120\,$Hz, $\langle r^E(t) \rangle_t=9.1$\,Hz and $\langle r^I(t) \rangle_t=14.1$\,Hz.

Once adapted and fitted, we calculate the net feedforward input to the (sup./int.) pyramidal cells and find that they are dominated by excitation (for all cells and all parameters).  
This strong drive that is net excitatory is crucial for determining which limit to take in the asymptotic calculation in Appendix A.  
In Appendix A (equation \eqref{qth_append}), by taking the limit of large voltage values, we show that the firing rate heterogeneity (std. dev.) is captured qualitatively by:
\begin{equation}\label{qth_rng_thry}
	\sigma\left(\vec{\nu}\right) \approx C \sqrt{ \frac{1}{N-1} \sum_{j=1}^N \left(\frac{q_j}{\theta_j} - \mu\left(\frac{\vec{q}}{\vec{\theta}}\right) \right)^2  }
\end{equation}
where $\mu\left(\frac{\vec{q}}{\vec{\theta}}\right)=\frac{1}{N}\displaystyle\sum_{j=1}^N \frac{q_j}{\theta_j}$.  
In the analysis, we focus on a specific limit of the {\it reduced} firing rate equations rather than dealing with a multidimensional partial differential equation.

\begin{figure}
	\centering
 	\includegraphics[width=\columnwidth]{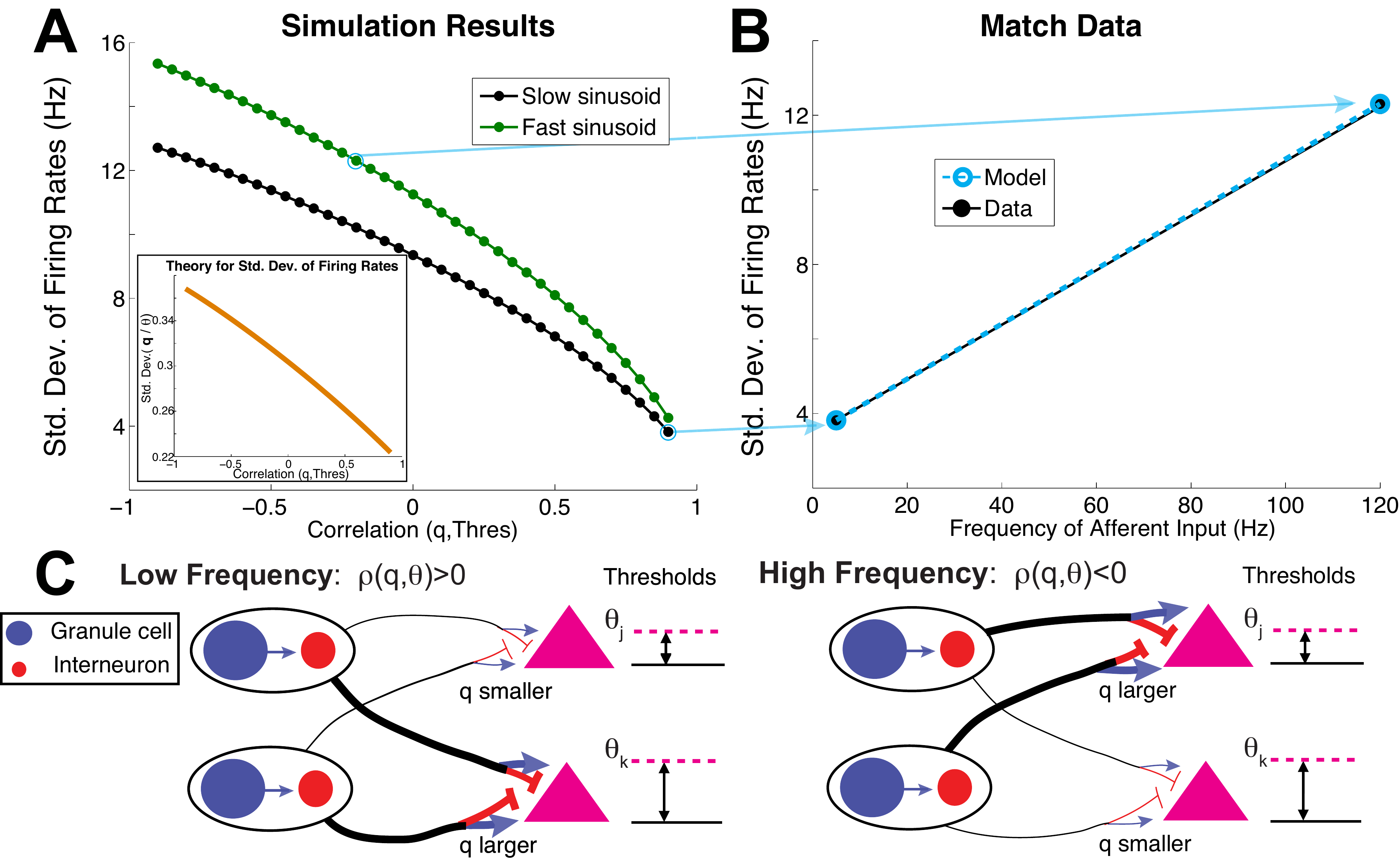}
\caption{Model prediction: the effective network input strength depends on the afferent stimuli and the excitability of the target neuron.  
{\bf A}) Simulations of the firing rate standard deviations in the delayed feedforward networks as the correlation between $\vec{q}$ and $\vec{\theta}$ vary with low (black) and high (green) frequency sinusoidal input.  
Inset: The standard deviation of firing rates as the correlation between $\vec{q}$ and $\vec{\theta}$ varies, using equation \eqref{qth_rng_thry}.  
{\bf B}) To capture the different firing rate standard deviations, consider different correlations of $(q,\theta)$ for each stimuli.  
In the data, with a low frequency (5\,Hz) stimuli, there were 15 neurons with trial averaged firing rate standard deviation of 3.8\,Hz (population average of 23.6\,Hz); 
with a high frequency (120\,Hz), there were 15 neurons with trial averaged firing rate standard deviation of 12.3\,Hz (population average of 23.6\,Hz).  In the model (cyan dots), with the same 
parameters as in Fig. \ref{fig:cycl} were used, but now the correlation $\varrho$ is allowed to vary.  For the low 
frequency, $\varrho=0.9$ gives the best approximation to the firing rate standard deviation of the data among the $\varrho$ mesh values considered.  For the high frequency, $\varrho=-0.2$ 
gives the best approximation to the firing rate standard deviation of the data.  
{\bf C}) Cartoon of a small sub network to illustrate how intrinsic threshold $\theta$ of 2 neurons is related to effective network input strength $q$ that consists of 
both excitatory and inhibitory inputs.  
For low frequencies, $\varrho(\theta,q)>0$ so that neurons with smaller $\theta$ tend to have smaller effective network input $q$ (and vice-versa).  
With high frequencies, $\varrho(\theta,q)<0$ so that neurons with smaller $\theta$ tend to have larger effective network input $q$ (and vice-versa).  
}
\label{fig:res2}       
\end{figure}

Figure \ref{fig:res2}{\bf A} demonstrates how the firing rate standard deviations 
change as the correlation $\varrho$ between threshold heterogeneity and synaptic variability vary.  When $\varrho<0$, larger $\theta_j$ tend to occur with smaller $q_j$, so the maximum of $q/\theta$ is amplified 
(big times big) and the minimum is diminished (small times small), resulting in relatively large heterogeneity.  When $\varrho>0$, larger $\theta_j$ tend to occur with larger $q_j$ so both the 
variance of $q/\theta$ is smaller (small times big minus big times small), resulting in a relatively small heterogeneity.

{\bf \underline{ Intuitive Understanding of Figure \ref{fig:res2}A.}}  A priori, guessing how the firing rate heterogeneity changes as the correlation of $\vec{\theta}$ and $\vec{q}$ varies is difficult and depends on models, regimes, etc. 
(see \citet{Ly_15} where the firing rate range depends nonlinearly on intrinsic and network heterogeneity).  
However, since the feedforward input is effectively excitatory in the models and regimes here, the results in Figure \ref{fig:res2}{\bf A} can be understood as follows.  If cells with high threshold have weak inputs (i.e., smaller $q$), they will have low rates; 
at the same time, cells with low thresholds have strong inputs (i.e., larger $q$), they will have high rates -- together, the whole population will have a broad distribution of rates.  
If, on the other hand, those cells with high thresholds receive strong inputs and those with low threshold receive weak inputs, then the distribution will be narrower.  The opposite trend would occur if the feedforward input is 
net inhibitory. 


Since the experimental data shows that the delayed feedforward input is crucial for observing {\bf less} firing rate heterogeneity with lower frequencies than with higher frequencies (Fig. \ref{fig:setup}{\bf B}), 
our model is applicable because the structure of this delayed feedforward input strongly effects the firing rates.  Indeed, the theory applied to these parameters (Fig. \ref{fig:res2}{\bf A}, inset) is validated in the 
large spiking network model (Fig. \ref{fig:res2}{\bf A}).  Here, the firing rate standard deviations are plotted while $\varrho(\vec{\theta},\vec{q})$ is varied while keeping the mean and variance (among the $N$ target cells) 
fixed.  There are two curves because two different sinusoidal frequencies $\phi$ were used: 5\,Hz in black and 120\,Hz in green.  
We remark that the statistics of the feedforward inputs remain the same throughout the various correlation values $\varrho(\vec{\theta},\vec{q})$, but since there is (colored) noise $\sigma_F\eta_l$, 
there are minor deviations for each simulated network due to finite simulations.


\subsection{Model prediction: strength of effective delayed feedforward inputs depends on stimuli}

A natural model prediction from these modeling results (Fig. \ref{fig:res2}{\bf B}) is that the effective delayed feedforward input strength is structured in a stimulus-dependent manner.  
With lower frequency afferent inputs, we use the model to predict that the pyramidal cells with higher thresholds (intrinsically less excitable) receive overall stronger delayed feedforward excitatory and inhibitory input than 
cells with lower thresholds (more excitable); that is, $\varrho>0$.  With higher frequency afferent inputs, the model predicts that pyramidal cells with higher threshold (intrinsically less excitable) receive overall 
weaker delayed feedforward excitatory and inhibitory input than cells with lower thresholds ($\varrho<0$).  

Figure \ref{fig:res2}{\bf B} shows the firing rate standard deviation from the experimental data (as a function of the dominant sinusoidal frequency), compared with the fitted model where we see different firing rate standard deviations 
depending on $\varrho$.  With the low frequency, we plot the firing rate standard deviation with the `best' corresponding $\varrho$ that is closest to the firing rate standard deviation from the data 
($\varrho=0.9$, recall the black curve in Fig. \ref{fig:res2}{\bf C}) in cyan.  With 
the high frequency, again we plot the firing rate rate with the `best' corresponding $\varrho$, which is $\varrho=-0.2$ (green curve in Fig. \ref{fig:res2}{\bf A}) in cyan.  
Given the good match between model and data, the application of the theory clearly demonstrates:  
\begin{itemize}
	\item The correlation $\varrho$ is significant in controlling the firing rate heterogeneity
	\item The specific network input strength (or structure) depends on both stimulus and targeted neuron
\end{itemize}
The difference in firing rate heterogeneity 
cannot be attributed to the sinusoidal frequency $\phi$ in this model because we have already seen that it does not vary much with the two different sinusoidal inputs (Fig. \ref{fig:res2}{\bf A}).  

Figure \ref{fig:res2}{\bf C} shows a schematic picture of the model prediction for two pyramidal cells with lower and higher thresholds receiving feedforward inputs.  
This prediction's viability relies on the network providing different sets of inputs for different stimulus frequencies -- a fact supported by previous experimental results \citep{bol11}.  
The thickness of the arrows indicate how strong the effective network input ($q$) is from both presynaptic E and I cells.  This model prediction is a statistical statement about the 
aggregate population and not about individual cells.  For example, the theory certainly allows 
for cells with higher thresholds to have larger effective network inputs with high frequency inputs (appearing to violate $\varrho<0$), but the correlation being negative (in this case) means these cases will happen 
less often than the other case.

\subsection{Linking theory to neural architecture}\label{sec:neurConn}

In the prior section we showed a direct application of the theory where the parameters $(\theta_j,q_j)$ were manipulated in the computational model, resulting in firing rate standard deviations 
captured by our analysis.  In this 
section, we apply the theory in a different way and link it with the architecture of a neural network.  A motivating reason for this is that many sensory systems can encounter both low and high frequency stimuli, 
and it is conceivable that it may need to process both low and high frequency stimuli in rapid succession.  Thus, the effective network input strength $\vec{q}$ may not be able to change fast enough, 
which would seemingly weaken our theory.  We present an alternative application of our theory using network coupling to address this.

\begin{figure}
	\includegraphics[width=\columnwidth]{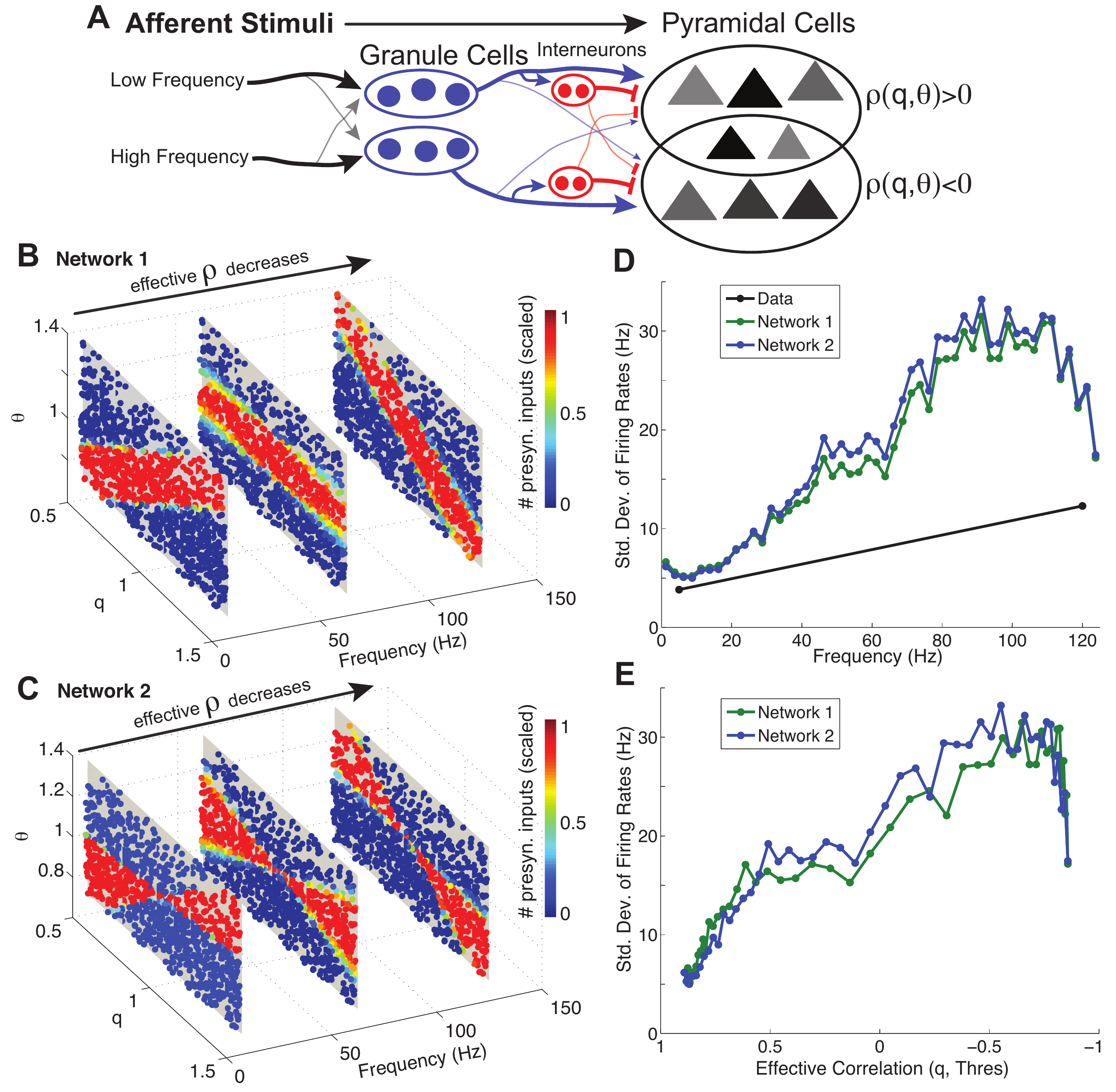}
\caption{Using the theory to capture the firing rates standard deviations in the data with fixed $(\vec{\theta},\vec{q})$, with $\varrho(\vec{\theta},\vec{q})=0$.  
{\bf A}) Demonstrates how the theory can be used to get different firing rates standard devaitions.  Depending on frequency on the afferent stimulus, selective activation of granule cells can 
provide effectively more presynaptic input to the {\bf desired} set of pyramidal cells that have a specific $\varrho(\vec{\theta},\vec{q})$ (i.e., $\varrho>0$ for low frequencies and $\varrho<0$ for 
higher frequencies).  
{\bf B}) Network 1: a particular network architecture that captures the desired $\varrho$, for fixed $(\vec{\theta},\vec{q})$, shown for three different frequencies.  The coloring indicates a measure 
of the effective input (red is more, blue is less): (number of presynaptic inputs from granule cells)$\times$(Probability granule cell fires), scaled to have values in [0,1].  
{\bf C}) Network 2: another network architecture that captures the desired $\varrho$.  
In both networks, we assume $\varrho$ varies continuously as the frequency increases; this assumption is not necessary and is only made to demonstrate the utility of the theory.  
{\bf D}) Comparing how the firing rate heterogeneity (measured by standard deviation) with frequency in both network models and in the experimental data.  Consistent with the theory, we see the heterogeneity generally increases 
with frequency.  In both network models, since the total numbers of granule and pyramidal cells are kept fixed (see Table \ref{tab:1}), the conductance values here are exactly the same as the previous figures.  
{\bf E}) The firing rate standard deviation as a function of the effective $\varrho$ ({\bf reverse scale}, positive is left, negative is right) from the two networks is consistent with the theoretical analysis.
}
\label{fig:archit}       
\end{figure}

The neural network model we consider here has {\it fixed} heterogeneity values $(\vec{\theta},\vec{q})$ for the target pyramidal cells, and we set $\varrho(\vec{\theta},\vec{q})=0$ (although we will see this is not necessary).  
The network is designed so that the {\it effective} correlation between $\vec{\theta}$ and $\vec{q}$ are as before (i.e., 
$\varrho>0$ for smaller $\phi$ and $\varrho<0$ for larger $\phi$).  
Networks in the previous sections have a 20\% connection probability, so some cells do not receive network inputs; by {\it effective} we mean that pyramidal cells that are actually connected the delayed feedforward have this $\varrho$.  
Here, only a subset of presynaptic granule cells will be activated and respond (i.e., spike) to sinusoidal input (previously they all responded equally), and different groups of these granule cells will respond 
{\it depending} on the sinusoidal frequency $\phi$ (see Fig. \ref{fig:sinTune}{\bf A} and Appendix B for details).  
Indeed, there is evidence based on recordings and theory for frequency tuning of granule cells in the ELL \citep{bol11}.  
By construction, pyramidal cells are differentially activated by delayed feedforward input, but note that as before, {\it all} pyramidal cells receive the same afferent sinusoidal input $I_{aff}$.  
The connectivity rules for the delayed feedforward network will no longer be random (Erd\H{o}s-R\'enyi graph, see Figure \ref{fig:sinTune}{\bf B}) 
but rather the \underline{activated presynaptic cells} have a higher probability of connecting to pyramidal cells 
that result in the desired $\varrho$; Figure \ref{fig:archit}{\bf A} shows a cartoon schematic of this idea.  In implementing this idea, we make the assumption that the 
effective $\varrho(\vec{\theta},\vec{q})$ varies continuously from a high (positive) value to low (negative) value as $\phi$ increases.  
There are numerous ways to effectively have various $\varrho$ values in the pyramidal cells, and we only present 2 instances of the network (see Appendix B 
for connectivity rules and other details).  Figure \ref{fig:archit}{\bf B} illustrates the connectivity rules for {\bf Network 1}: the activated presynaptic cells at a low frequency have a higher probability of connection to the 
red cells on a diagonal band in $(\vec{q},\vec{\theta})$ space (color bar represents actual number of inputs with various $\phi$).  The reason for this diagonal band with positive slope is that the effective $\varrho$ will be 
positive.  As $\phi$ increases, other sets of presynaptic cells are activated and the probability of connection is again higher for target cells that are in the red band.  
Figure \ref{fig:archit}{\bf C} illustrates the connectivity rules for {\bf Network 2} that are similar to the previous network except the region is different.  
The activated presynaptic cells at a low frequency have a higher probability of connection to the 
red cells that form two triangular wedges in $(\vec{q},\vec{\theta})$ space.  Again, as $\phi$ increases, other sets of presynaptic cells are activated and the probability of connection is higher for cells that are in the red band.  

In these two networks ({\bf Network 1, 2}), there are overall less presynaptic cells activated than before (we kept the same $N_f=100$ and $N=1000$), but even keeping 
the same conductance strengths $(s_e, s_i)$ happened to result in firing rates that are comparable to the previous model (Fig. \ref{fig:res2}).  In general, the 
conductance strengths $(s_e,s_i)$ may need to change to account for different presynaptic cell activation.  
With a fixed architecture, afferent sinusoidal input is provided to all populations as before and only the frequency $\phi$ is varied 
(with selective activation of presynaptic granule cells).  We simply linearly interpolate the values of the amplitude $\mathcal{A}$ and mean $I_0$ sinusoidal input between 5\,Hz and 120\,Hz 
(recall the slightly different values in Table \ref{tab:1}).  
The resulting firing rate standard deviations (Fig. \ref{fig:archit}{\bf D}) are again described by our reduced analytic descriptions (Eq. \eqref{qth_rng_thry}).  
Specifically, the effective 
correlation $\varrho$, which in this case was attributed to a particular frequency $\phi$, is the determining factor in the firing rate standard deviation.  Figure \ref{fig:archit}{\bf E} shows this directly; the firing 
rate standard deviations from these two networks are plotted as a function of effective correlation\footnote{In Fig. \ref{fig:archit}{\bf E}, 
the effective correlation for a given $\phi$ is the Pearson's correlation calculated on the set of $(\theta_j,q_j)$ weighted by the number of presynaptic inputs.} 
(reverse scale on x-axis) and again the standard deviation is larger with $\varrho<0$ than $\varrho>0$.  
There is a drop in firing rate range in Fig. \ref{fig:archit}{\bf E} around -1 due to it being a hard bound on correlation; but there is also a drop in Fig. \ref{fig:archit}{\bf D} after $\sim100$\,Hz following a relatively steady (yet noisy) increase in the firing rate 
range in both networks. This is likely because the scheme we chose to illustrate this principle did not insure that all frequencies received the same total sinusoidal drive, together with the structured connectivity rules resulted in different ranges of inputs 
(across the population) that varied with frequency. It is difficult to insure strict monotonicity in the firing rate range because of all the interacting components, yet our theory is valuable in understanding how this complicated statistic modulates.  
Unlike the previous neural network, we were able to obtain 
different firing rate standard deviations with the same fixed $(\vec{\theta},\vec{q})$ and fixed connectivity that is structured as opposed to completely random.

We chose to illustrate our result with two networks to show that although there are differences in the actual range of firing rates that depend on numerous factors (network connectivity rules, input parameters, etc.), 
the qualitative results of our theory holds and thus we expect the results to hold with other prescribed network configurations.

Our model and the results of Figure \ref{fig:archit} demonstrate the value of our theoretical results.  These theoretical results not only enable a specific prediction about how the effective feedforward 
inputs are related to excitability, but can also be applied to predict probability of connections based on intrinsic excitability and frequency tuning.

\section{Discussion}

We have adapted and applied a theory for firing rate heterogeneity in recurrent networks to a delayed feedforward network, and used data from the electrosensory system of the weakly electric fish to 
motivate this theoretical application and constrain the model.  Our data shows that firing rate heterogeneity is larger for higher frequency sinusoidal input 
(courtship of opposite sex) than with lower frequency sinusoidal input 
(antagonizing signal with same sex animals), and that this difference in heterogeneity is determined by the delayed feedforward.  Our theory and computational models predict that, in order to 
account for the observed firing rate statistics, the effective network connectivity is dynamically modulated depending on stimulus features.  
Our work uses theoretical analysis to specifically predict how the interactions of neural attributes (threshold heterogeneity and synaptic strengths) lead to heterogeneous firing rate statistics with various stimuli.  These 
results may be generalizable to other feedforward neural networks.

Our theoretical analysis replicates a fairly simple configuration were the correlation between synaptic strength and threshold can influence population heterogeneity. 
Our analysis reveals that this relationship is non-trivial in that it is a nonlinear function of synaptic strength divided by threshold that controls the firing rate heterogeneity across the population rather than how these individual quantities alter a cell's firing rate 
(see Fig. \ref{fig:fr_indiv} in Appendix C for how these individual quantities are related to firing rate).

Other cellular or network features certainly effect the firing rate heterogeneity, but we focus on threshold heterogeneity because threshold is known to be important 
in this system \citep{middleton09}, and other cellular attributes can be related to 
the threshold in the leaky integrate-and-fire model \citep{mejias12,mejias14}.  Besides the previously mentioned reasons for the existence and importance of different synaptic strengths, 
our data shows that delayed feedforward input activity from the specific network of granule cells significantly affects the firing rate distribution because when it is blocked, the firing rate 
heterogeneity does not vary much.  
Several other attributes of sensory neurons, and the distribution of their heterogeneity will influence the way they encode information (tuning, non-linearity, plasticity, to name only a few).   
Investigating how network dynamic influences these heterogeneities is a rich research direction for future studies.

Although the fitted model was best described by the `rhythmic' regime where the presynaptic inputs were large, the theory in \citet{Ly_15} also accounts for asynchronous regimes where the 
presynaptic inputs are weaker and 
background fluctuations significantly affects spiking.  In those regimes, the relationship between the heterogeneous attributes lead to different firing rate heterogeneity than the `rhythmic' regimes.  
Thus, the experimental predictions can be reexamined and potentially augmented if other fitted models happen to operate in a different regime.  

Whether our theoretical predictions are verified in the weakly electric fish are yet to be determined, however 
we discuss various possibilities here.  An alternative explanation for the modulation of firing rate heterogeneity is that the 
heterogeneity of the presynaptic firing rate could modulate with stimulus (in our models the presynaptic firing rates are statistically homogeneous) and the superficial pyramidal cells would inherit this 
heterogeneity.  However, this would not explain all of the experimental data: with the afferent input strength being identical in all cases, blocking the delayed feedforward input increased the firing rate 
heterogeneity for low frequencies (and decreased it for high frequencies). Clearly, the delayed feedforward input cannot simply contribute to the cells' firing rate heterogeneity, 
otherwise blocking the feedback would systematically reduce heterogeneity by removing a source of variability. Instead, our model shows that firing rate 
heterogeneity can either increase or decrease simply but varying the correlation between network input strength and the cells' excitability (threshold). 
Experimental verification that this mechanism does indeed play a role in the ELL would require a 
thorough set of experiments, probably involving both in-vitro and in-vivo recording. 
We know that this mechanism is not the only one affecting firing rate heterogeneity in the ELL. For example, a previous study showed that the delayed feedforward input interacts 
with the burst-generating mechanism to produced stereotyped responses (less heterogeneous) during low-frequency stimulation \citep{marsat2012prep}. 
This study can, in fact, fit the theory presented here. If one considers bursting as a source of 
intrinsic excitability, that paper showed strong positive network input interacts with this intrinsic excitability, the implication being that a positive correlation between the two would lead to strong stereotyped 
bursts (low heterogeneity). 
Clearly, ELL pyramidal cells have many more properties than are being explicitly modeled here, but it is precisely the reason that makes our theory so general and applicable to a wide variety of systems.

We focus specifically on firing rate heterogeneity because it is the first order statistic and crucial in understanding the mechanisms that lead to efficient coding.  
The effects of heterogeneity of neural attributes on coding and dynamics of neural networks is important and has been studied by numerous authors both in a general theoretical framework 
\citep{hermann12,mejias12,hunsberger14,mejias14} and in specific neural systems \citep{shamir06,chelaru08,marsat10,marsat12}.  Our results differ from these studies in many ways, but 
largely because i) we account for two heterogeneous attributes, ii) we make a specific prediction about synaptic strengths depending on the pyramidal (target) cells and stimulus features.  
The results here may be applicable to other systems given how common 
feedforward pathways are and how well spiking neuron models can capture the statistics of neural network activity.

Firing rate heterogeneity is just one statistical measure of the network response, and there are other measures that may also be crucial in the context of neural coding.  The spike coherence of the response with 
the afferent stimuli is commonly used in the electrosensory system \citep{chacron05,mehaffey08,middleton09}, and thus an important future direction is how the heterogeneity of the firing rate in time affects the efficiency of coding.  In addition to the afferent sinusoidal input, communication signals in electric fish also consist of brief chirps.  However, the predominant 
electrical signal in time is the nearly sinusoidal input that we have considered.  The afferent sinusoidal 
activity without the transient chirp at the very least sets the stage for the neural network to readily decode the chirp input, and is thus an important component \citep{marsat2012prep}.  
Another potentially important 
measure is the second order statistics, or the spike count correlation, of the pyramidal cells \citep{averbeck06,cohen11,doiron16}.  
A few studies have recently considered how spike count correlation in the ELL might effect coding of signals: \citet{litwin12} considered how local/global stimulation alters correlation at 
different time windows of superficial pyramidal cells, \citet{simmonds15} considered the how noise (and signal) correlation modulates with granule cell (parallel fiber) input, and \citet{chacron08} showed how 
the noise correlation of these neurons depends on stimulus type and firing pattern (bursts).  
Theoretical analyses of such measures when focusing on heterogeneity are complicated \citep{josic09,ecker11,LMD_whisker_12} and beyond the scope of this current paper.

\begin{acknowledgements}
	We thank David J. Edwards (VCU) for help regarding Power analysis of our data.  
	This work was supported by a grant from the Simons Foundation (\#355173, Cheng Ly), and by a grant from the National Science Foundation (NSF-ISO \#1557846, Gary Marsat).
\end{acknowledgements}

\section*{Appendix A: Asymptotic calculation for the relative standard deviation of firing rate distribution}\label{thry_q_theta}

The asymptotic calculations were all based on an expression for the firing rate of an individual neuron via the 
PDF or Population Density framework that has been commonly used in spiking models in cortex models \citep{knight72b,wilburrinzel82,fourcaudbrunel,brunel_latham,tranchinaBookChapt} and other areas 
\citep{barna,brownmoehlisholmes,huertassmith}.  In addition to the firing rate, this framework has been useful in calculating many statistical quantities of the spike train \citep{BCFA,richardson07,richardson08} and to study the stability of 
coupled networks \citep{knight72b,abbottvv,brunelhakim,gerstner00,lyErm_siads_10}.  It can also be employed as a time saving computational tool \citep{ntfast,oks,at06,ly_tranchina_07}.  We focus on the standard deviation 
of firing rates and use the framework to gain analytic insight into the dynamics.

To employ the framework and for feasibility, we make some technical assumptions:
\begin{enumerate}[(i)]
	\item the (average) population firing rate is a good approximation to the presynaptic input rate with random connectivity
	\item a single p.d.f. function is adequate to describe a single population's activity
	\item the heterogeneity is driven by $(q_j,\theta_j)$ only
\end{enumerate}
The complexities that arise in a recurrent network (nonlinear PDF equation) never came about \citep{Ly_15} in our specific analysis because of the networks here are delayed feedforward networks.  Moreover, 
the resulting PDF equations have lower dimensions than a recurrent network.  We begin by writing down the probability density function for both (E and I) presynaptic populations that 
provide delayed feedforward input.  
$$ \rho^{E}(v,\eta,t) \,dv d\eta = \Pr\left(\hbox{presyn E cell}\in\{(v,v+dv)\cap(\eta,\eta+d\eta)\} \right) $$
$$ \rho^{I}(v,\eta,t) \,dv d\eta = \Pr\left(\hbox{presyn I cell}\in\{(v,v+dv)\cap(\eta,\eta+d\eta)\} \right) $$
The PDF equations for the presynaptic E population are:
\begin{eqnarray}
	\frac{\partial \rho^E}{\partial t} &=& -\frac{\partial}{\partial v}\left\{ J_V(v,\eta,t) \right\} 
	-\frac{\partial}{\partial \eta}\left\{J_\eta(v,\eta,t) \right\} \nonumber \\
	J_V(v,\eta,t) &=& \frac{1}{\tau_m} \left[  I_{aff}(t)-v + \sigma_F \eta \right]\rho^E   \\
	J_\eta(v,\eta,t) &=& \frac{1}{\tau_n}\left[-\eta\rho^E - \frac{1}{2}\frac{\partial\rho^E}{\partial \eta} \right]   \\
	J_V(1,\eta,t) &=& J_V(0,\eta,t+\tau_{fref}) \\
	r^E(t) &=& \int_{-\infty}^\infty J_V(1,\eta,t)\,d\eta  
\end{eqnarray}
where $r^E(t)$ is the population firing rate.  The presynaptic I population is similar but lacks direct sinusoidal input:
\begin{eqnarray}
	\frac{\partial \rho^I}{\partial t} &=& -\frac{\partial}{\partial v}\left\{ J_V(v,\eta,t) \right\} 
	-\frac{\partial}{\partial \eta}\left\{J_\eta(v,\eta,t) \right\} \nonumber \\
	J_V(v,\eta,t) &=& \frac{1}{\tau_m} \left[  -v - g_{EE}(t)(v-\mathcal{E}_E) + \sigma_F \eta \right]\rho^I   \\
	J_\eta(v,\eta,t) &=& \frac{1}{\tau_n}\left[-\eta\rho^I - \frac{1}{2}\frac{\partial\rho^I}{\partial \eta} \right]   \\
	J_V(1,\eta,t) &=& J_V(0,\eta,t+\tau_{fref}) \\
	r^I(t) &=& \int_{-\infty}^\infty J_V(1,\eta,t)\,d\eta	 
\end{eqnarray}

For simplicity, we assume the synaptic conductances of the target (superficial pyramidal) cells can be averaged in time (see equation \eqref{f_lif}): 
$$ g_{EE}(t) = s_{EE} \int  r^E(t-t') K(t') \,dt' $$
$$ g_e(t) = s_e \int  r^E(t-t') K(t') \,dt' $$
$$ g_i(t) = s_i \int  r^I(t-t') K(t') \,dt' $$
where $K$ is the alpha function kernel:
$$ K(t) = H(t)\frac{\alpha}{\frac{\tau_d}{\tau_r}-1}\Big[  e^{-t/\tau_d} - e^{-t/\tau_r} \Big]$$
and $H(t)$ is the Heaviside step function and $\tau_r<\tau_d$ (a common assumption in models of synapses).  This kernel is unconventional in that $\int_{-\infty}^\infty K(t)\,dt=\alpha\tau_r$ and $\neq1$.

Finally, the PDF for the target cells: 
$$ \rho(v,\eta,t) \,dv d\eta = \Pr\left(\hbox{pyramidal cell}\in\{(v,v+dv)\cap(\eta,\eta+d\eta)\} \right) $$
are described by:
\begin{eqnarray}
	\frac{\partial \rho}{\partial t} &=& -\frac{\partial}{\partial v}\left\{ J_V(v,\eta,t) \right\} 
	-\frac{\partial}{\partial \eta}\left\{J_\eta(v,\eta,t) \right\} \nonumber \\
	J_V(v,\eta,t) &=& \frac{1}{\tau_m} \Big[  I_{aff}(t)-v - q_j g_e(t-\tau_{del})(v-\mathcal{E}_E) -q_j g_i(t-\tau_{del})(v-\mathcal{E}_I) + \sigma_P \eta \Big]\rho   \\
	J_\eta(v,\eta,t) &=& \frac{1}{\tau_n}\left[-\eta\rho - \frac{1}{2} \frac{\partial\rho}{\partial \eta} \right]   \\
	J_V(1,\eta,t) &=& J_V(0,\eta,t+\tau_{ref}) \\
	r_j(t) &=& \int_{-\infty}^\infty J_V(\theta_j,\eta,t)\,d\eta. 
\end{eqnarray}
The firing rate $r_j(t)$ is not a population firing rate, but rather the firing rate of the $j^{th}$ neuron in the population.  We implicitly assume that the only difference between cells is given 
by the two heterogeneous parameters: $(\theta_j,q_j)$.  

Our goal is not to capture the time-varying firing rates (which are still difficult even with these assumptions because of the three coupled PDEs that each have 2 spatial dimensions and time), but rather 
we are interested in the trial- and time-averaged firing rates.  This enables a compact expression for how the heterogeneity ($\vec{\theta},\vec{q}$) relationship controls the heterogeneity of steady-state firing rates.  We have:
\begin{equation}
	\nu_j := \langle r_j(t) \rangle_t  
\end{equation}
We approximate 
\begin{equation}\label{tmavg_approx}
 	\Big\langle \int_{-\infty}^\infty I_{aff}(t)\rho(v,\eta,t)\,d\eta \Big\rangle_t  \approx  \int_{-\infty}^\infty \langle I_{aff}(t) \rangle_t \langle \rho(v,\eta,t)\rangle_t \,d\eta,
\end{equation}
and similarly for the expressions with $g_{e/i}$.  This leads to:
\begin{equation}
	\nu_j \approx \frac{1}{\tau_m} \Big[ I_0 - \theta_j + \bar{g}_E (\mathcal{E}_E-\theta_j) - \bar{g}_I (\theta_j-\mathcal{E}_I) \Big] f(\theta_j) + \frac{\sigma_P}{\tau_m} \int_{-\infty}^\infty \eta \rho(\theta_j,\eta)\,d\eta
\end{equation}
where $f(v)=\int \rho(v,\eta)\,d\eta$ is the steady-state marginal voltage distribution (equal to the time-average, assuming ergodic theorems apply), 
and $\bar{g}_E=\alpha\tau_r \langle r^E(t)\rangle_t$, $\bar{g}_I=\alpha\tau_r \langle r^I(t)\rangle_t$.  One could simply numerically simulate these equations, but there is not much 
analytic insight gained in understanding how ($\vec{\theta},\vec{q}$) and $\varrho(\vec{\theta},\vec{q})$ alter the firing rate standard deviation.  In applying dimension reduction methods, there 
are issues that arise in trying to accurately capture the firing rate \citep{ly_tranchina_07}.  Thus, we apply a simple (quantitatively inaccurate) dimension reduction method 
where we assume $\eta$ is frozen and average over the resulting firing rate \citep{moreno3,nesse08,hertag14,nlc_15,Ly_15}.  We also ignore the effects of the refractory period 
$\tau_{ref}$ \footnote{Although ignoring the refractory period could be problematic for large firing rates, we emphasize that the purpose of our this analysis is not for quantitative matching but rather for an analytic 
explanation.  A similar calculation with the refractory period was performed (not shown) but the results were not insightful.}.  The firing rate is then simply:
\begin{eqnarray}
\nu_j(\theta_j,q_j) &\approx& \int_{-\infty}^\infty \nu_{det}(\theta_j,q_j; \eta) \frac{e^{-\eta^2}}{\sqrt{\pi}} \,d\eta  \label{nu_integral} \\
	 \nu_{det}(\theta_j,q_j; \eta)  &=& \begin{cases} 
	0, & \text{if } \frac{q(\bar{g}_E\mathcal{E}_E+\bar{g}_I\mathcal{E}_I)+\sigma_P\eta+I_0}{1+q(\bar{g}_E+\bar{g}_I)} \leq  \theta_j \\
	\frac{1+q(\bar{g}_E+\bar{g}_I)}{\tau_m
	\log\left(\frac{q(\bar{g}_E\mathcal{E}_E+\bar{g}_I\mathcal{E}_I)+\sigma_P\eta+I_0}
		{q(\bar{g}_E\mathcal{E}_E+\bar{g}_I\mathcal{E}_I)+\sigma_P\eta+I_0-\theta(1+q(\bar{g}_E+\bar{g}_I))}\right)},  & \text{if }
		\frac{q(\bar{g}_E\mathcal{E}_E+\bar{g}_I\mathcal{E}_I)+\sigma_P\eta+I_0}{1+q(\bar{g}_E+\bar{g}_I)} >  \theta_j 
	\end{cases} \label{nu_aprx}
\end{eqnarray}

The parameters $(\theta_j,q_j)$ determine how one $\nu_j$ differs from another; to see how the combined effects of threshold heterogeneity and synaptic variability alter $\nu_j$, we consider a specific limit.  
That is, the simulations indicate that the the net conductance are large in the fitted model (Fig. \ref{fig:chooseThry}), thus, we consider the large firing rate limit of the term in the integrand $\nu_{det}$, to get:
\begin{eqnarray}
	\tau_m \nu_{det}(\theta_j,q_j) &=& \frac{1+q_j(\bar{g}_E+\bar{g}_I)}{\log\left(\frac{q_j(\bar{g}_E\mathcal{E}_E+\bar{g}_I\mathcal{E}_I)+\sigma_p\eta+I_0}
		{q_j(\bar{g}_E\mathcal{E}+\bar{g}_I\mathcal{E}_I) +\sigma_p\eta+I_0 - \theta_j(1+q_j(\bar{g}_E+\bar{g}_I))} \right)} \\
		&=& \frac{q_j}{\theta_j}(\bar{g}_E\mathcal{E}_E+\bar{g}_I\mathcal{E}_I)+\frac{\sigma_p\eta+I_0}{\theta_j}-\frac{1}{2}(1+q_j(\bar{g}_E+\bar{g}_I)) \nonumber \\
		& &-\frac{(1+q_j(\bar{g}_E+\bar{g}_I))^2\theta_j}{12[q_j(\bar{g}_E\mathcal{E}_E+\bar{g}_I\mathcal{E}_I)+\sigma_p\eta+I_0-\theta_j(1+q_j(\bar{g}_E+\bar{g}_I))]} \nonumber \\
		& &+O\left(z^2(1+q_j(\bar{g}_E+\bar{g}_I))\right) \label{rhyth_expan} \label{asymp_exp} \\
	 \hbox{where } z &:=& \theta_j \frac{1+q_j(\bar{g}_E+\bar{g}_I)}{q_j(\bar{g}_E\mathcal{E}_E+\bar{g}_I\mathcal{E}_I)+\sigma_p\eta+I_0-\theta_j(1+q_j(\bar{g}_E+\bar{g}_I))} \label{z_defn}
\end{eqnarray}
This calculation is very similar to the one in \cite{Ly_15}.  The key term is the first term in equation \eqref{asymp_exp}, 
$$ \frac{q_j}{\theta_j}(\bar{g}_E\mathcal{E}_E+\bar{g}_I\mathcal{E}_I) $$
which is the dominant term assuming $\nu_{det}$ is large.  Substituting the expansion \eqref{asymp_exp} into the integral approximation \eqref{nu_integral} only changes terms with $\eta$ in them (i.e., the 
dominant term does not change and the term $\sigma_P\eta/\theta_j$ evaluates to 0).  
This shows analytically that the term $q_j/\theta_j$ is the dominant source of firing rate heterogeneity, and that we can approximate:
\begin{equation}\label{qth_append}
	\sigma\left(\vec{\nu}\right) \approx C \sqrt{ \frac{1}{N-1} \sum_{j=1}^N \left(\frac{q_j}{\theta_j} - \mu\left(\frac{\vec{q}}{\vec{\theta}}\right) \right)^2  }
\end{equation}
where $\mu\left(\frac{\vec{q}}{\vec{\theta}}\right)=\frac{1}{N}\displaystyle\sum_{j=1}^N \frac{q_j}{\theta_j}$.

\section*{Appendix B: Details of network connectivity and model in section \ref{sec:neurConn}}

We used two networks, which we generically labeled as  {\bf Network 1} and {\bf Network 2} (see Fig. \ref{fig:archit}), 
to help demonstrate the utility of the theory for firing rate standard deviations with different architectures than random (Erd\H{o}s-R\'enyi graph).  
We first describe how we selectively activate the granule cells that provide delayed feedforward input to the pyramidal cells, and then describe the connectivity rules for each network.

\begin{figure}
\centering
 	\includegraphics[width=0.5\columnwidth]{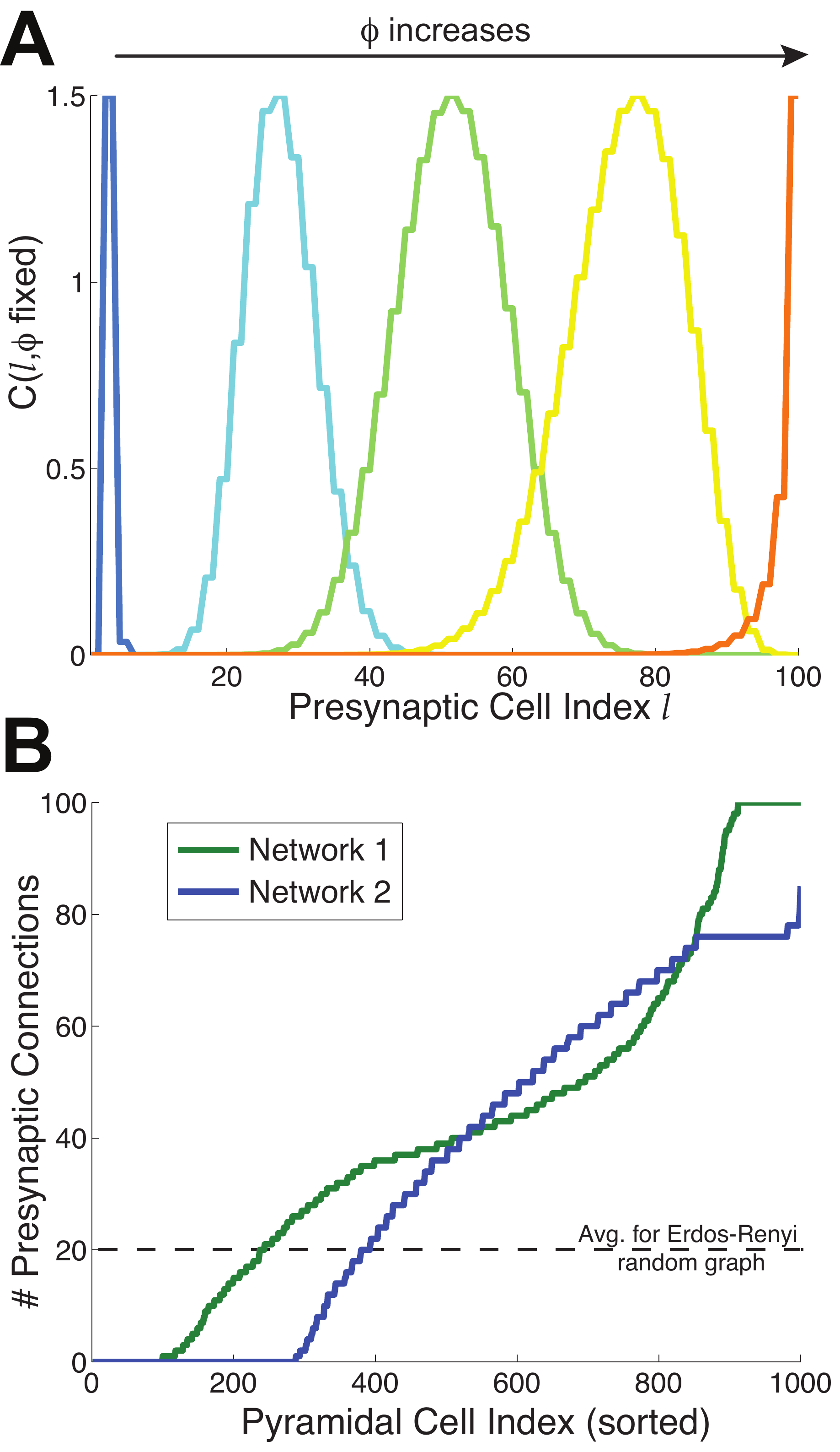}
\caption{Activation of presynaptic cells depends on the frequency of the afferent sinusoidal input.  
{\bf A}) The strength of the sinusoidal input $C(l,\phi)$ (see equation \eqref{sin_strng}) for particular frequencies; from left to right, $\phi=3.75, 33.75, 63.75, 93.75, 123.75$\,Hz.  
{\bf B}) Showing the number of presynaptic connections (out of $N_f=100$) for each of the $N=1000$ target pyramidal cells (sorted from smallest to largest) in both networks.  Notice the 
structure of the connectivity, compared to completely random connectivity (black dash line, with 20\% connection probability).
}
\label{fig:sinTune}       
\end{figure}

\underline{Selective activation of granule cells.}
Instead of providing constant sinusoidal drive to each of the $2N_f$ presynaptic granule cells, the afferent stimuli $I_{aff}(t)=I_0+\mathcal{A}\sin(2\pi\phi t)$ is scaled by a parameter $C(l,\phi)$:
$$ I_{aff}(t) = C(l,\phi) \Big\lceil I_0 + \mathcal{A}\sin(2\pi\phi t) \Big \rceil^+$$
that depends on both frequency $\phi$ and the index of cell $l\in\{1,2,\dots,N_f\}$ ($2N_f$ total because there are both excitatory and inhibitory presynaptic cells).  
The afferent stimuli to the target pyramidal cells is the same as before (see equation \eqref{e_lif}).  
Before providing the formula for $C(l,\phi)$, we note that the strength of the sinusoidal drive will follow a (scaled) beta distribution where the location of the maximum value increases as $\phi$ increases.  
We use:
\begin{equation}\label{sin_strng}
	C(l,\phi) = 1.5\frac{x(l)^{19}(1-x(l))^{21.52*\frac{125}{\phi} -21}}{\displaystyle\max_{1\leq l \leq N_f} \beta(l,\phi)}
\end{equation}
where 
\begin{eqnarray}
	\beta(l,\phi) &=& x(l)^{19}(1-x(l))^{21.52*\frac{125}{\phi} -21}\nonumber \\
	x(l) &=& \frac{2.5\,\hbox{Hz}}{125\,\hbox{Hz}}(\lceil l/4 \rceil - 1/2) \nonumber
\end{eqnarray}
The term $\beta(l,\phi)$ is simply the numerator of the fraction in $C(l,\phi)$ so that $C(l,\phi)\in(0,1.5)$.  The variable $x(l)$ is the end result of mapping the $l^{th}$ presynaptic neuron to one of 50 frequencies 
(equally spaced by 2.5\,Hz from 1.25\,Hz to 123.75\,Hz) and normalizing by 125\,Hz.  
Here we are assuming $\phi\in[0,125]$\,Hz so that $x(l)\in(0,1)$; other frequencies can easily be incorporated with minor adjustments to the above formulas.  Finally, notice that in $x(l)$, we have the term: 
$\lceil l/4 \rceil$, which denotes rounding up after dividing by 4; this essentially groups presynaptic neurons into groups of size 4 that receive the same sinusoidal drive.  Figure \ref{fig:sinTune}{\bf A} illustrates how the 
sinusoidal drive to the presynaptic cells, indexed by $l$, vary with several fixed frequencies $\phi$.

\underline{Connectivity rules for Network 1.}
To illustrate the usefulness of the theory,  we implemented a static delayed feedforward network with fixed $(\vec{\theta},\vec{q})$ and certain connectivity rules (see below).  The presynaptic granule cells are indexed 
as before via $l$, and the target pyramidal cells by $j$.  Recall that both $E$ and $I$ cells in the presynaptic population have the {\it same} connectivity for simplicity.  

Each pyramidal cell $j$ has an associated $(\theta_j,q_j)$ (Fig. \ref{fig:archit}{\bf B},{\bf C}), and since each presynaptic $l$ cell's sinusoidal drive depends on frequency, the probability of connection is specified so that 
the effective $\varrho(\theta,q)$ results in firing rate heterogeneity consistent with the data (i.e., $\varrho>0$ for low frequencies and $\varrho<0$ for high frequencies).  The connection 
probability is closely related to Figure \ref{fig:archit}{\bf B}; that is: 
{\it Low frequencies activate a subset of presynaptic cells, the probability that those presynaptic cells are connected to target cells with $(\theta_j,q_j)$ in a region that gives $\varrho>0$ is high 
(Fig. \ref{fig:archit}{\bf B}, red); the probability is lower with the other target cells (blue).}  Similar connection probability rules apply for high frequencies.  In both networks we considered, the connectivity scheme assumes that 
as the afferent sinusoidal frequency increases, $\varrho$ decreases monotonically.  Again, this is a questionable assumption that we have made only to provide a proof of principle for how our theory can be used.  
In \underline{Network 1}, different effective $\varrho$ values are obtained by lines with slopes proportional to $\varrho$.  Of course, there are an infinite number of ways to arrive at a desired $\varrho$ value.  The probability of a connection is:
\begin{eqnarray}
	P(l\hbox{ is connected to }j) &=& e^{-100 d(l,j)} \nonumber \\
	\hbox{where }d(l,j) &=& \min_{(\theta_0,q_0)\in\mathbb{B}(l)}  \sqrt{ (\theta_j-\theta_0)^2+(q_j-q_0)^2 } \nonumber \\
	 \mathbb{B}(l) &=& \Big\{ (\theta,q) \Big\vert m_l(q-\bar{q})-0.1 < \theta-\bar{\theta} < m_l(q-\bar{q})+0.1  \Big\} \nonumber \\
	 m_l = \big(1-2x(l) \big)\frac{\displaystyle\max_j \theta_j - \displaystyle\min_j \theta_j}{\displaystyle\max_j q_j - \displaystyle\min_j q_j}; 
	& \hspace{.2in} &\bar{q}=\frac{1}{2}\left( \displaystyle\max_j q_j + \displaystyle\min_j q_j \right);  
	 \hspace{.2in} \bar{\theta}=\frac{1}{2}\left( \displaystyle\max_j \theta_j + \displaystyle\min_j \theta_j \right); \nonumber 
\end{eqnarray}
$x(l)\in(0,1)$ was defined above (i.e., scaled frequency that drives $l$ well).  
The function $d(l,j)$ is the Euclidean distance in $(\theta,q)$ space to the band $\mathbb{B}(l)$, which consists of $(\theta,q)$ values that give the desired 
$\varrho$.  Note that the slope of the band $m_l$ goes from positive to negative as $l$ increases.

\underline{Connectivity rules for Network 2.}
The rules in this network are similar in spirit to \underline{Network 1} but the region that gives an effective $\varrho$ value is no longer rectangular but rather a wedge (compare Fig. \ref{fig:archit}{\bf B} and {\bf C}).  
The probability of a connection is:
\begin{eqnarray}
	P(l\hbox{ is connected to }j) &=& e^{-100 d(l,j)} \nonumber \\
	\hbox{where }d(l,j) &=& \min_{(\theta_0,q_0)\in\mathbb{W}(l)}  \sqrt{ (\theta_j-\theta_0)^2+(q_j-q_0)^2 } \nonumber \\
	 \mathbb{W}(l) &=&  \mathbb{W}_1(l) \cup \mathbb{W}_2(l) \nonumber \\
	 \mathbb{W}_1(l) &=& \Big\{ (\theta,q) \Big\vert (n_l - 0.4)(q-\bar{q}) < \theta-\bar{\theta} < (n_l+0.4)(q-\bar{q})  \Big\} \nonumber \\
	 \mathbb{W}_2(l) &=& \Big\{ (\theta,q) \Big\vert (n_l + 0.4)(q-\bar{q}) < \theta-\bar{\theta} < (n_l - 0.4)(q-\bar{q})  \Big\} \nonumber \\
	 n_l = \big(0.8-1.6x(l) \big)\frac{\displaystyle\max_j \theta_j - \displaystyle\min_j \theta_j}{\displaystyle\max_j q_j - \displaystyle\min_j q_j}; 
	& \hspace{.2in} &\bar{q}=\frac{1}{2}\left( \displaystyle\max_j q_j + \displaystyle\min_j q_j \right);  
	 \hspace{.2in} \bar{\theta}=\frac{1}{2}\left( \displaystyle\max_j \theta_j + \displaystyle\min_j \theta_j \right); \nonumber 
\end{eqnarray}
$x(l)\in(0,1)$ has the same definition as before, but the function $d(l,j)$ is now the Euclidean distance in $(\theta,q)$ space to the region $\mathbb{W}(l)$ which are 2 
triangular wedges (Fig. \ref{fig:archit}{\bf C}).

The resulting number of connections for both {\bf Network 1, 2} are not random but rather has structure (Fig. \ref{fig:sinTune}{\bf B}).  

\section*{Appendix C: More Data and Model Figures }

%
%

Here we provide supplemental figures for completeness; we chose not to include them in the main manuscript for exposition purposes.

\begin{figure}
\centering
 	\includegraphics[width=\columnwidth]{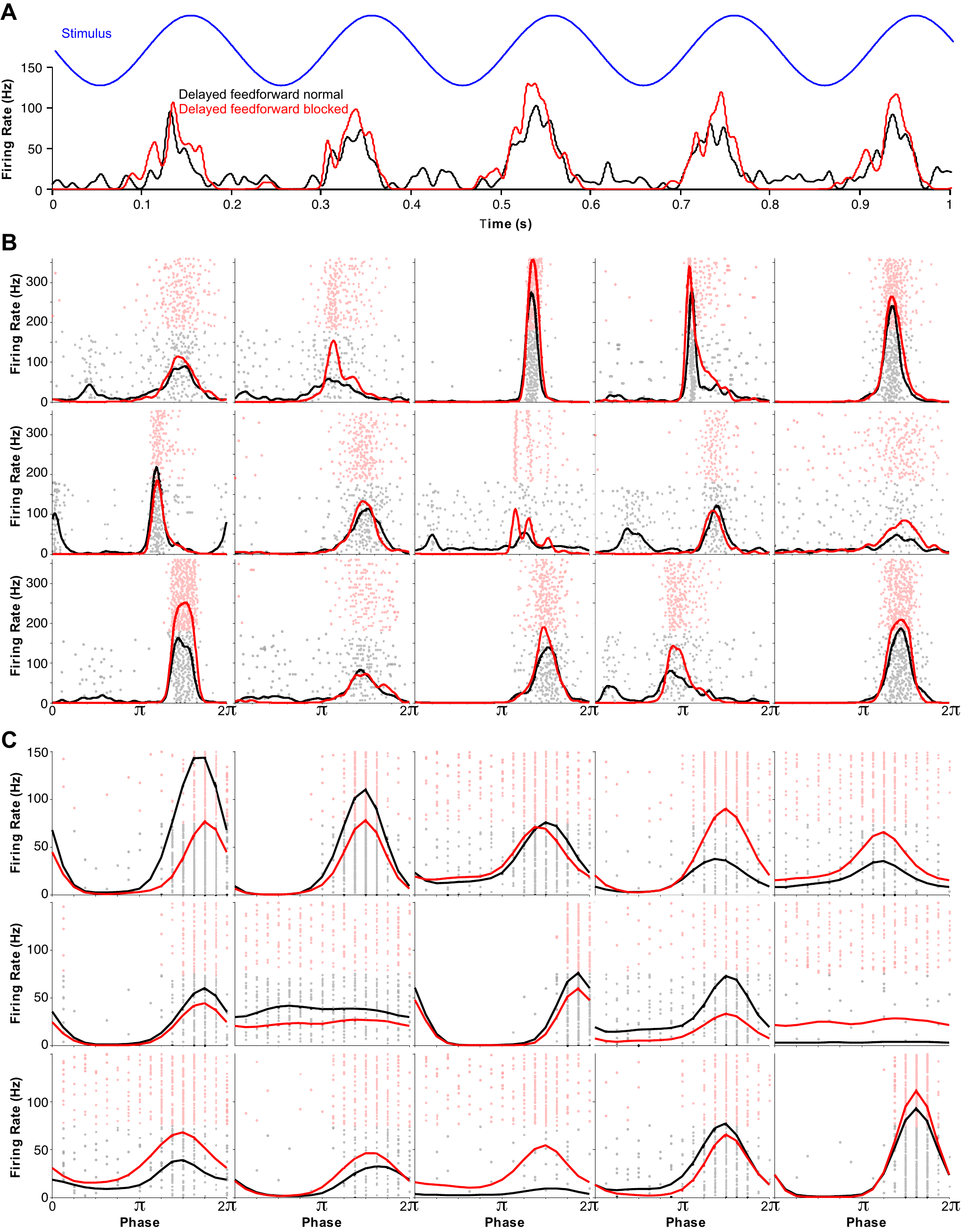}
\caption{
Responses of the 15 pyramidal cells in our data set. Parallel fiber input was intact (black) or blocked (red).  {\bf A}) One second excerpt of the responses to the 5\,Hz AM stimulus (blue).  
Responses were converted to instantaneous firing rates (i.e. convolved with a 10\,ms Gaussian) and averaged across the 15 cells of our data set.  
{\bf B}) PSTH (cycle histogram) or responses of each cell to 5\,Hz stimuli.  Raster plots of spiking for each cycle are shown in the background. C) Same as B but with high frequency input (120\,Hz).  
}
\label{fig:allCyc}       
\end{figure}

Figure \ref{fig:allCyc} shows the entire experimental data set used in this paper.  Figure \ref{fig:allCyc}{\bf A} shows the 5\,Hz sinewave stimulus (blue), and the PSTH averaged over all 15 superficial/intermediate 
pyramidal cells in the intact network (black) and with parallel fiber input blocked (red).  Parts {\bf B} and {\bf C} show the individual cell PSTH (cycle histogram) for both intact (black) and blocked (red) conditions with 
solid lines; the raster plots for all trials are shown in the background (same format as Figure \ref{fig:setup}{\bf B}).

\begin{figure}
\centering
 	\includegraphics[width=0.85\columnwidth]{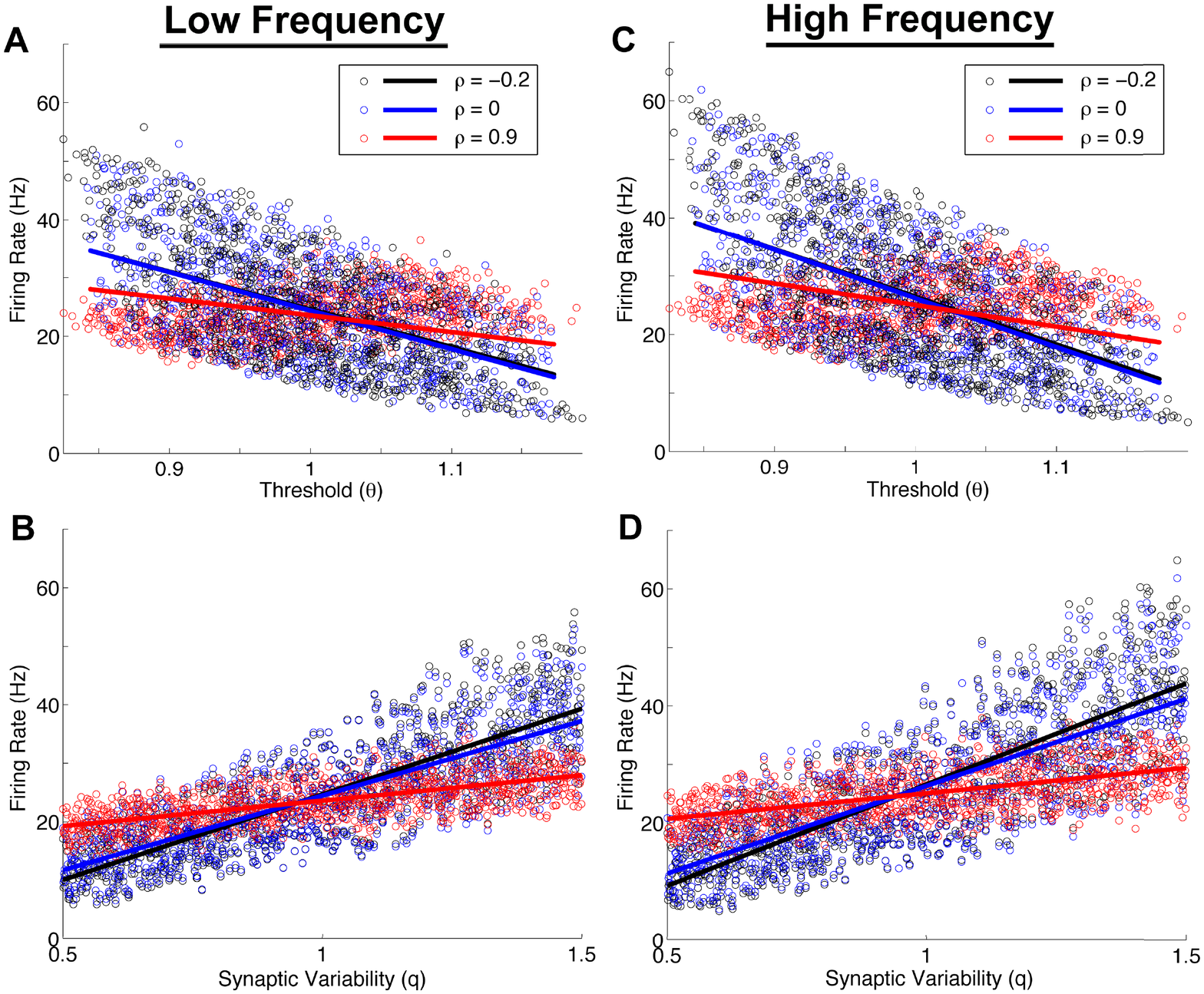}
\caption{Resulting firing rate as a function of heterogeneous parameter values.  
{\bf A}) Relationship between firing rate and threshold $\theta$ for all 1000 LIF neurons with representative correlation $\varrho(q,\theta)\in\{-0.2,0,0.9\}$ values for $\phi=5$\,Hz. 
{\bf B}) Relationship between firing rate and $q$ for all 1000 LIF neurons with representative correlation $\varrho$ values for $\phi=5$\,Hz.  
{\bf C}) and {\bf D}) similar to {\bf A}) and {\bf B}) but with $\phi=120$\,Hz input frequency. Although the firing rate is strongly related to the threshold, the synaptic variability has a weaker 
although positive correlation with the resulting firing rate.}
\label{fig:fr_indiv}       
\end{figure}

Figure \ref{fig:fr_indiv} shows the firing rate of the entire population of $N=1000$ (target) cells as a function of the chosen thresholds and synaptic variability parameters, for both low and high frequency stimuli.  
Each panel contains relevant correlation values $\varrho(q,\theta)$.

\bibliographystyle{spbasic}      


\end{document}